\documentclass[reprint,amsmath,amssymb,aps,]{revtex4-1}
\usepackage{natbib}
\setcitestyle{square,numbers}
\usepackage{graphicx}
\usepackage{dcolumn}
\usepackage{bm}
\usepackage{amsmath}
\usepackage[utf8]{inputenc}
\usepackage{balance}
\usepackage{multirow}
\usepackage[caption=false]{subfig}
\usepackage{longtable}
\usepackage{hyperref}

\begin{document}


\title{Exploring the accurate nuclear potential}

\author{D.K. Swami, Yash Kumar}
\author{T. Nandi}%
\affiliation{Inter University Accelerator Centre, New Delhi-110067}

\date{\today}

\begin{abstract}
We have constructed empirical formulae for fusion and interaction barrier heights using experimental values available in the literature. Fusion excitation function measurements are used for the former and back angle quasi-elastic excitation function for the latter case. The fusion barriers so obtained have been compared with various model predictions such as Bass potential, Christenson and Winther, Broglia and Winther, Aage Winther, Siwek-Wilczyńska and J.Wilczyński, Skyrme energy density function model, and the Sao Paulo optical potential along with experimental results. The comparison allows us to find the best model, which is found to be the Broglia and Winther model. Further, to examine its predictability, the Broglia and Winther model parameters are used to obtain total fusion cross sections showing good agreement with the experimental values for beam energies above the fusion barriers. Thus, this model can be useful for planning any experiments, especially ones aiming for super heavy elements. Similarly, current interaction barrier heights have also been compared with the Bass potential model predictions. It shows that the present model calculations are much lower than the Bass potential model predictions. We believe the current interaction barrier model prediction will be a good starting point for future quasi-elastic scattering experiments. Whereas both the Broglia and Winther model and our interaction barrier model will have practical implications in carrying out physics research near the Coulomb barrier energies.\\
\begin{description}
\item[PACS numbers] 24.10.-i, 25.70.Jj, 25.70.Bc, 21.30.Fe, 25.55.Ce
\end{description}
\end{abstract}

\maketitle


\section{Introduction}

The basic characteristics of nuclear reactions are usually described by an interaction consisting of a repulsive Coulomb potential term and a short ranged attractive nuclear potential term. The resultant potential can be expressed as a function of the distance between the centres-of-mass of the target and projectile nuclei. When a projectile approaches the target nucleus it experiences a maximum force at a certain distance where the repulsive and attractive forces cannot balance each other, the repulsive Coulomb force is always higher. The projectile needs to overcome the barrier for coming close to the target nucleus. This barrier is referred to the fusion barrier $(B_{fu})$, which is a basic parameter in describing the nuclear fusion reactions. The kinetic energy of the projectile must be adequate to surmount this barrier in order to enter a pocket at the adjacent to the barrier at a shorter distance, where the nuclei undergo nuclear fusion processes. The barrier is determined by the excitation function measurement of nuclear fusion \cite{BB}, whereas it is estimated by many theoretical models such as Bass potential model \cite{Bass, bass}, proximity potential model \cite{Blocki}, double folding model \cite{Satchler}, Woods-Saxon potential model \cite{Saxon}, and semi-empirical models such as Christenson and Winther (CW) model \cite{Christensen}, Broglia and Winther (BW) model \cite{Reisdorff}, Aage Winther (AW) model \cite{AW}, Denisov potential (DP) model \cite{Dutt}, Siwek-Wilczyńska and Wilczyński (SWW) model \cite{Siwek}, Skyrme potential model \cite{Zanganeh}, and the Sao Paulo optical potential (SPP)\cite{Freitas}.

The quasi-elastic (QEL) processes involving relatively small energy transfer to excite only nuclear levels in either one of the participating nuclei or in both become significant as soon as the two bodies approach within the range of the nuclear forces. The position where the resultant of the Coulomb and nuclear forces is still repulsive and additional energy is required to get the two bodies interacting over their mutual potential barrier. This barrier is hitherto somewhat smaller than the fusion barrier and is known as the interaction barrier $ (B_{int}) $, which is measured by the excitation function studies of QEL scattering experiment \cite{Mitsuoka}. Obviously, the two barriers are different from each other as one is characterized by the fusion reaction and the other by the QEL scatterings \cite{Bass,bass}. However, many-a-time the distinction is overlooked, for example,  \cite{Mitsuoka,dutt}. Worth noting here that it is only the Bass model \cite{Bass,bass} which can estimate the interaction barriers in addition to the fusion barriers.

Recently Sharma and Nandi \cite{Prashant} demonstrated the coexistence of the atomic and nuclear phenomenon on the elastically scattered projectile ions while approaching the Coulomb barrier. Here the projectile ion x-ray energies were measured as a function of ion beam energies for three systems $^{12}$C($^{56}$Fe,$^{56}$Fe), $^{12}$C($^{58}$Ni,$^{58}$Ni) and $^{12}$C($^{63}$Cu,$^{63}$Cu) and observed unusual resonance like structures as the beam energy approaching the fusion barrier energy according to the Bass model \cite{Bass, bass}. We expected the resonance near to interaction barrier as this technique resembled the quasi-elastic (QEL) scattering experiment \cite{Mitsuoka}. To resolve this anomaly, we planned to examine both the fusion and interaction barriers in greater detail.

At present many experiments exist in the literature for the measurements of the fusion and interaction barrier heights. In this work, we have constructed an empirical formula for estimating the fusion barriers from the fusion excitation function measurements alone and another for the interaction barriers from the QEL scattering experiments only. In the next step, we have compared the empirical results so obtained with the predictions from the various models based on proximity type of potentials such as Bass potential \cite{Bass, bass} and Christenson and Winther (CW) \cite{Christensen} and Woods-Saxon type of potentials such as Broglia and Winther (BW) \cite{Reisdorff}, Aage Winther (AW) \cite{AW}, Siwek-Wilczyńska and Wilczyński (SWW) \cite{Siwek}, Skyrme Potential (SP) \cite{Zanganeh} models, and the Sao Paulo optical potential (SPP) \cite{Freitas}. This comparison suggests for the need for further measurements in certain specified regions for both the cases. It is seen that this work will be useful in various applications \cite{BB}, for example, correct prediction of $B_{fu}$ may play important roles in experiments for the formation of super heavy elements \cite{Tathagat} and that of $B_{int}$ finds its significance in physics research near the Coulomb barriers \cite{Prashant}.


\section{Determination of the barrier heights}
The fusion cross section is plotted as a function of the beam energy to obtain the excitation function curve for the corresponding reaction. Whereas the fusion barrier is obtained from the barrier distribution plot, which is defined as the second derivative of the energy-weighted cross section $\frac{d^2(\sigma E)}{dE^2}$ versus beam energy in the center-of-mass frame \cite{Rowley}. Many articles report only the excitation function, in such cases the fusion cross section and corresponding beam energy have been multiplied to plot against the beam energy to obtain the double derivative. The $\frac{d^2(\sigma E)}{dE^2}$ plot against beam energy have been fitted with the Gaussian function to obtain the fusion barrier heights.\\

Similarly, the interaction barrier can be obtained from the QEL excitation function studies as follows. The QEL scattering is affected by the sum of elastic, inelastic, and transfer processes, which is measured at backward angles of nearly $180^0$, where the head-on collisions are dominant. The barrier distribution is obtained by taking the first derivative, with respect to the beam energy, of the QEL cross-section relative to the Rutherford cross section, that is, $\frac{-d}{dE} (\frac{d\sigma_{QEL}}{d\sigma_R})$ \cite{Andres}. This method has been examined in several intermediate-mass systems \cite{Timers,Hagino}. One can notice that the QEL barrier distribution behaves similarly to the fusion barrier distribution, although the former is less sensitive to the nuclear structure effects. 

\section{General background}
 Theoretically, the total nucleus-nucleus interaction potential $V_T(R)$ between the projectile and  target nuclei, in general, is given by\\
\begin{equation}
V_T(R)=V_N (R) + \frac{l(l+1)\hbar^2}{2\mu R^2} + V_c (R) 
\label{eq1}
\end{equation}
Where the first term $V_N(R)$ is the model dependent nuclear potential, the second is centrifugal potential, where $\mu = \frac{m_p m_t}{m_p+m_t}$ is the reduced mass of the projectile mass $ m_p$ and the target nuclei mass $m_t$ in MeV/$c^2$ units and $l$ represents the angular momentum of the two body system. When we consider the fusion barrier of the system, $l$ is set to zero which means the centrifugal or the second term is zero. The third term is the Coulomb potential similar to \cite{Birkelund} and given by \\
\begin{equation}
 V_c(R) = \frac{Z_1 Z_2 e^2}{4\pi\epsilon_0}
  \begin{cases}
    \frac{1}{R}       & \quad \text{for } R \geq R_B\\
    \frac{1}{2R_B} \Big[3-\Big(\frac{R_B}{R}\Big)^2\Big]  & \quad \text{for} R<R_B
  \end{cases}
   \label{eq2}
\end{equation}
Where the fusion barrier radius $R_B$ = $R_c$($A_p^\frac{1}{3}$ + $A_t^\frac{1}{3}$), $R_c$ depends on the system as discussed below and $A_p$ and $A_t$ are mass of
projectile and target nuclei, respectively. Putting the first term of equation (\ref{eq1}) from any particular model, one can get the fusion barrier radius, $R_B$ by using the condition\\
\begin{equation}
\frac{dV_T(R)}{dR}\bigg|_{R=R_B} = 0
 \label{eq3}
\end{equation}
and 
\begin{equation} 
\frac{d^2V_T(R)}{dR^2}\bigg|_{R=R_B} \leq 0
 \label{eq4}
  \end{equation}
  Hence, $V_T(R=R_B)$ = $B_{fu}$ and similarly, $V_T(R=R_{int})$ = $B_{int}$, where $R_{int}$ is the interaction barrier radius. Of course, $V_N(R)$ in equation (\ref{eq1}) can be replaced by appropriate one, for the interaction barrier case as done in Bass potential model \cite{Bass}. \\
  
\section{Nuclear potential models}
 We shall briefly discuss different models for $V_N(R)$ including one developed from fully experimental results in the present work.
 
 \subsection{Present empirical model}
According to the definition for the Coulomb potential given above, the shape of the nuclear potential discussed below and representation of nuclear distances, for example, one given above for the barrier radius, $B_{fu}$ and $B_{int}$ may be written as a function of $Z_p$, $Z_t$, $A_p$ and $A_t$. Hence, the experimentally obtained $B_{fu}$ from fusion excitation function measurements and $B_{int}$ from QEL measurements can be plotted against $z$, where $z = \frac{Z_p Z_t}{(A_p^\frac{1}{3}+A_t^\frac{1}{3} )}$, as shown in Fig.1(a) and Fig.1(b), respectively. The fusion experiments used for Fig. 1 (a) and the QEL experiments used for Fig. 1(b) are given in Table I and II, respectively. Fusion data are available for $8\leq z \leq278$, whereas QEL data for $59\leq z \leq313$. We can notice that both $B_{fu}$ vs $z$ and $B_{int}$ vs $z$ are nonlinear. The whole range of data can either be fitted with two straight lines or a third-degree polynomial to obtain the reduced chi-square nearly equal to one. Nevertheless, the later fitting is found to be somewhat better and is thus used in this work.

\begin{table}[h!]
\centering
\caption{The fusion barriers $(B_{fu})$ for the following two body systems have been used in Fig 1(a) to obtain the empirical formula for estimating the $B_{fu}$ for any system in the bound $8\leq z \leq278$.}
\begin{tabular}{ p{4cm}p{3cm}p{1.3cm}  }
\hline
\hline
System & $z$ & $B_{fu} (MeV)$   \\      \hline \\
$^{12}$C+$^{15}$N & 8.83 &  6.80 \cite{Louis C. Vaz}\\              
$^{12}$C+$^{16}$O & 9.98 & 7.50 \cite{sperr}\\ 
$^{12}$C+$^{26}$Mg & 13.71& 11.5 \cite{C. M. Jachcinski}\\ 
$^{12}$C+$^{30}$Si & 15.57 & 13.2 \cite{L. C. Vaz}\\
$^{16}$O+$^{27}$Al & 18.84 & 43.6   \cite{Eisen}\\ 
$^{24}$Mg+$^{26}$Mg & 24.63 & 20.8 \cite{S. Gary}\\
$^{26}$Mg+$^{32}$S & 31.28 & 27.5 \cite{G. M. Berkowitz}\\
$^{12}$C+$^{92}$Zr & 35.27 & 32.3 \cite{JR Newton}\\
$^{16}$O+$^{72}$Ge & 38.32 & 35.4 \cite{E. F. Aguilera}\\
$^{32}$S+$^{40}$Ca & 48.52 & 43.3 \cite{V. Zanganeh}\\
$^{48}$Ca+$^{48}$Ca & 55.03 & 51.7 \cite{V. Zanganeh}\\
$^{27}$Al+$^{70}$Ge & 58.42 & 55.1 \cite{V. Zanganeh}\\
$^{32}$S+$^{58}$Ni & 63.58 & 59.5  \cite{Gutbrod}\\ 
$^{40}$Ar+$^{58}$Ni & 69.13 & 66.32\cite{V. Zanganeh}\\
$^{37}$Cl+$^{73}$Ge & 72.42 & 69.20\cite{V. Zanganeh}\\
$^{40}$Ca+$^{62}$Ni & 75.90 & 72.3 \cite{L. C. Vaz} \\
$^{32}$S+$^{89}$Y & 81.68 & 77.8 \cite{A. Mukherjee}\\
$^{16}$O+$^{238}$U & 84.43 & 80.8 \cite{V. Zanganeh}\\
$^{28}$Si+$^{120}$Sn & 87.84 & 85.9 \cite{V. Zanganeh}\\
$^{48}$Ca+$^{96}$Zr & 97.41 & 95.9\cite{A. M. stefanini}\\
$^{40}$Ca+$^{96}$Zr & 100.01 & 93.6 \cite{Gutbrod}\\ 
$^{40}$Ar+$^{121}$Sb & 109.72 & 111 \cite{Gauvin}\\ 
$^{40}$Ca+$^{124}$Sn & 118.95 & 113 \cite{Scarlassara}\\ 
$^{28}$Si+$^{198}$Pt & 123.18 & 121  \cite{Nishio}\\
$^{40}$Ar+$^{154}$Sm & 127.11 & 121 \cite{Reisdorf}\\ 
$^{40}$Ar+$^{165}$Ho & 135.43 & 141.4\cite{V. Zanganeh}\\
$^{40}$Ca+$^{192}$Os & 165.42 & 168.1\cite{V. Zanganeh}\\
$^{84}$Kr+$^{116}$Cd &186.67 & 204  \cite{gauvin}\\
$^{74}$Ge+$^{232}$Th & 278.45 & 310  \cite{gauvin}\\
\hline
\hline
\end{tabular}
\end{table}

\begin{table}[h!]
\centering
\caption{The interaction barriers $(B_{int})$ for the following two body systems have been used in Fig 1 (b) to obtain the empirical formula for estimating the $B_{int}$ for any unknown systems in the range $59\leq z \leq313$.}
\begin{tabular}{p{4cm}p{3cm}p{1.3cm}}
\hline
\hline
System & $z$ & $B_{int (MeV)}$   \\ \hline  \\   
$^{12}$C+$^{205}$Ti & 59.37 & 56.0\cite{Beyec}\\ 
$^{12}$C+$^{209}$Bi & 60.55 & 57.0\cite{Beyec}\\ 
$^{12}$C+$^{238}$U & 65.04 & 62.2\cite{Viola}\\ 
$^{14}$N+$^{238}$U & 74.82 & 73.4\cite{Viola}\\ 
$^{16}$O+$^{205}$Ti & 76.99 & 77.0\cite{Beyec}\\ 
$^{16}$O+$^{238}$U & 84.43 & 82.5\cite{Viola}\\ 
$^{20}$Ne+$^{238}$U & 103.23 & 102\cite{Viola}\\ 
$^{40}$Ar+$^{164}$Dy & 133.57 & 135\cite{beyec}\\ 
$^{40}$Ar+$^{238}$U  & 172.19 & 171\cite{bass}\\ 
$^{48}$Ti+$^{208}$Pb & 188.72 & 190.1\cite{Mitsuoka}\\
$^{54}$Cr+$^{208}$Pb & 202.78 & 205.8\cite{Mitsuoka}\\
$^{56}$Fe+$^{208}$Pb & 218.65 &223\cite{Mitsuoka}\\
$^{58}$Ni+$^{208}$Pb & 231.33 & 236\cite{Mitsuoka}\\
$^{70}$Zn+$^{208}$Pb & 244.86 & 250.6\cite{Mitsuoka}\\
$^{84}$Kr+$^{232}$Th & 307.86 & 332\cite{bass}\\ 
$^{84}$Kr+$^{238}$U  & 313.14 & 333\cite{Moretto}\\ 
\hline
\hline
\end{tabular}
\end{table}

\begin{figure}[h]
\begin{tabular}{@{}c@{}}
\includegraphics[width=\linewidth]{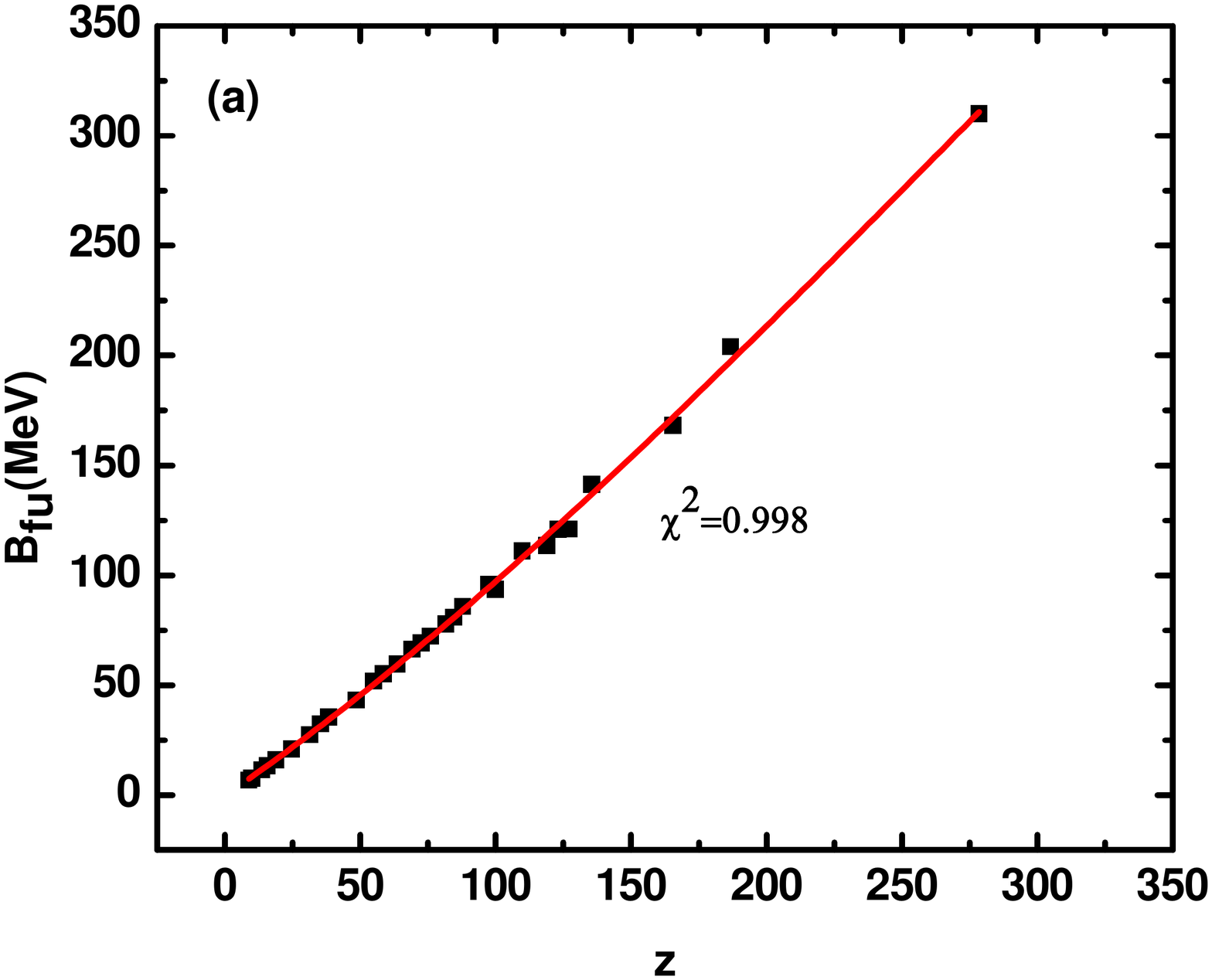}

\label{fig:subim1}
\end{tabular}
\begin{tabular}{@{}c@{}}
\includegraphics[width=\linewidth]{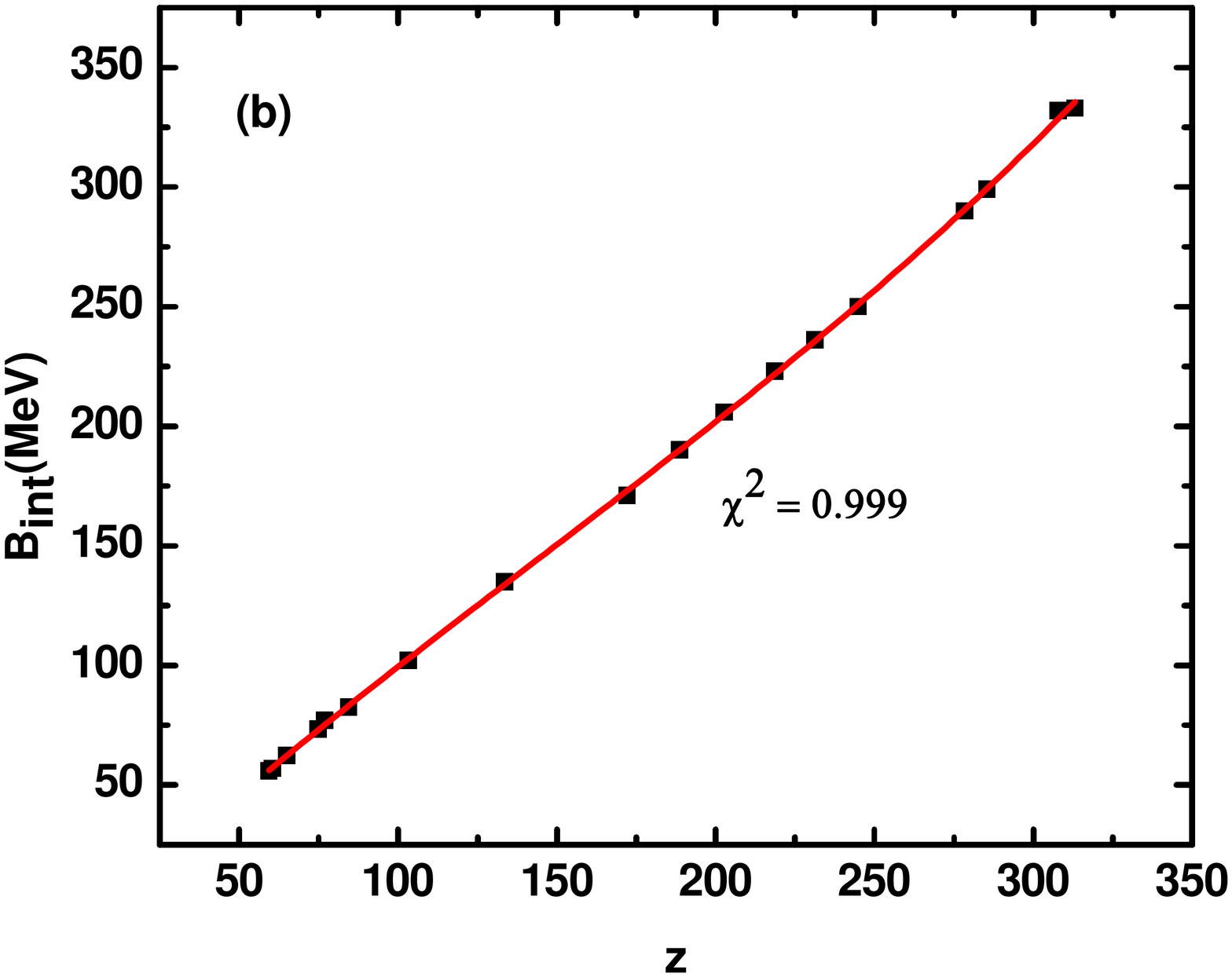} 

\label{fig:subim2}
\end{tabular}

\caption{Fusion and interaction barrier plots: (a) the fusion barrier, $ B_{fu}$ (MeV), as obtained from the fusion excitation function experiments and (b) the interaction barrier, $B_{int}$ (MeV), from the quasi-elastic scattering experiments, are plotted against the dimensionless parameter $z$ and the plots have been fitted with polynomial functions.}
\label{fig:image2}
\end{figure}

The polynomial function that fits the fusion excitation function data is as follows:\\
\begin{center}
$B_{fu}=-0.22+(0.85)z+(0.001)z^2-(1.75\times10^{-6})z^3$
\end{center}
\begin{equation}  
\begin{split}
8 \leq z  \leq 278
\end{split}
 \label{eq5}
\end{equation}
and the other function that fits the $B_{int}$ vs $z$ data is given below:
\begin{center}
$B_{int}=-13.47+(1.26)z-(0.002)z^2+(4.03\times10^{-6})z^3$
\end{center}
\begin{equation}
\begin{split}
59 \leq z \leq 313          
\end{split}
 \label{eq6}
\end{equation}
The empirical formula given in equation (\ref{eq5}) can predict $B_{fu}$ for any system within $8 \leq z  \leq 278$. Such predictions have been compared with different models. Similarly, the $B_{int}$ values obtained from equation (\ref{eq6}) we collated with the predictions of the Bass model.\\

To obtain $R_B$, we follow the following method.
The barrier height can be given as
\begin{equation}
    V_B(R) = \frac{Z_p Z_t e^2}{4\pi\epsilon_0 R} + V_N(R)
     \label{eq7}
\end{equation}
and at $R=R_B$
\begin{equation}
    V_B(R=R_B) = B_{fu}=\frac{Z_p Z_t e^2}{4\pi\epsilon_0 R_c(A_p^\frac{1}{3}+A_t^\frac{1}{3})} + V_N(R_B)
    \label{eq8}
\end{equation}
This equation can be written in terms of $z$
\begin{equation}
   B_{fu} = \frac{e^2}{4\pi\epsilon_0 R_c} z + V_N(R_B) 
   \label{eq9}
\end{equation}
For instance, if we assume $V_N(R)$ in the form of Woods-Saxon potential and it is not clearly a function of $z$, assumed to be a constant. Hence, the derivative of the above equation w.r.t. $z$ is as follows.
\begin{equation}
    \frac{dB_{fu}}{dz} = \frac{e^2}{4\pi\epsilon_0 R_c}
    \label{eq10}
    \end{equation}
    Now we replace $\frac{dB_{fu}}{dz}$ from equation (\ref{eq5}) we get
    \begin{equation}
      R_c = \frac{e^2}{4\pi\epsilon_0 (0.85 + (0.002)z - (5.25\times10^{-6}){z^2})}
        \label{eq11}
    \end{equation}
    and thus the barrier radius $R_B$.
  \subsection{Bass potential model}
  Bass potential model \cite{Bass,bass} suggests that the Coulomb barrier for quasi-elastic surface reaction is in general different from the Coulomb barrier for fusion. The former results from the quasi-elastic processes, where no mass or energy transfer takes place, whereas both mass and energy can transfer in the latter. Further, the quasi-elastic processes become significant as the projectile and target nuclei approach to the range of nuclear forces, where the resultant of the Coulomb and nuclear forces is still repulsive. Thus additional energy is required to get the nuclei within the resultant attractive force, where the fusion can occur. According to the Bass model, the barriers responsible in the quasi-elastic reactions is called the interaction barrier $B_{int}$ and it can be determined from the elastic scattering experiments \cite{Mitsuoka}. The other barrier is significant in fusion reactions and is defined as the fusion barrier $B_{fu}$. The latter is equal to the height of the potential barrier for zero angular momentum.\\
  The total effective Bass potential consists of a Coulomb, nuclear and centrifugal terms,
  \begin{equation}
 V_l (r)= \frac{Z_p Z_t e^2}{4\pi\epsilon_0 r} + \frac{\hbar^2 l^2}{2\mu r^2} - a_s  A_p^\frac{1}{3}  A_t^\frac{1}{3} \frac{d}{R}  {e^{-(\frac{r-R}{d})}}
 \label{eq12}
  \end{equation}
  Where $d$ is the range of nuclear interaction. The influence of fragment (projectile or target nuclei) properties on this potential can be expressed in terms of the dimensionless parameters 
  \begin{equation}
  x = \frac{e^2}{r_0 a_s4\pi\epsilon_0}   \frac{Z_p Z_t}{A_p^\frac{1}{3} A_t^\frac{1}{3} (A_p^\frac{1}{3} + A_t^\frac{1}{3} )}
     \label{eq13}
    \end{equation}
    \begin{equation}
    y= \frac{\hbar^2}{2m_0 r_0^2 a_s} \frac{A_p+A_t}{A_p^\frac{1}{3} A_t^\frac{1}{3} (A_p^\frac{1}{3} + A_t^\frac{1}{3} )}
     \label{eq14}
    \end{equation}
    Where $x$ is the ratio of the Coulomb force to the nuclear force and $yl^2$ is the ratio of the centrifugal force to the nuclear force at the point of contact i.e.,  $R_{pt} = r_0 (A_p^\frac{1}{3}+A_t^\frac{1}{3} )$, $r_0$=1.07 fm. Here,  $a_s$ =17.23 MeV is the surface constant as used in the liquid drop model of fission, $m_0$ the mass of a nucleon, $A_p$ and $A_t$ the mass number of projectile and target nuclei, respectively and other notations have usual significance.  Whereas $B_{fu}$ acts at $r$ = $R_{pt}$ + $d_{fu}$, $d_{fu}$ is the fusion distance. The $B_{int}$ is applicable for an interaction distance between the two surfaces $(d_{int})$ or the centre to centre distance, $R_{int}$ = $R_{pt}$ + $d_{int}$. The $d_{int}$ is always longer than the $d_{fu}$.\\
    
     The fusion distance $d_{fu}$ can be approximately obtained from the relation
     \begin{equation}
   \frac {d_{fu}}{d} \approx -\frac{lnx}{(1- \frac{2d}{R} )}
         \label{eq15}
         \end{equation}
The $d_{fu}$ varies with the fragments in the nuclear interactions. The barriers $B_{fu}$ and $B_{int}$ can be obtained from \\
    \begin{equation}
 B_{fu} = \frac{Z_p Z_t e^2}{4\pi\epsilon_0 R}  \Big\{\frac{R}{R+d_{fu}} - \frac{1}{x} \frac{d}{R}   e^{\big(-\frac{d_{fu}}{d}\big)}\Big\}
     \label{eq16}
    \end{equation}
    \begin{equation}
  B_{int} =\frac{Z_p Z_t e^2}{R+d_{int}}- 2.90 \frac{A_p^\frac{1}{3} A_t^\frac{1}{3}}{(A_p^\frac{1}{3}+A_t^\frac{1}{3})}
      \label{eq17}
      \end{equation}
      \begin{equation}
      d_{int}=2d = 2 \times 1.35=2.70 fm
           \label{eq18}
           \end{equation}
           Here, it is assumed that $d$ is independent of the mass of the nuclei. Hence,  from the above relations, we get $B_{fu} > B_{int}$.\\
           
  \subsection{Christenson and Winther model}
  Christenson and Winther \cite{Christensen} derived the nucleus-nucleus interaction potential on the basis of semi classical arguments as given by
  \begin{equation}
   V_N^{CW}(R) = - 50 {\overline{R}}{ e^{\big(-\frac{R-R_{pt}}{a}\big)}} MeV\\
   \label{eq19}
   \end{equation}
   where $R_{pt}$ = $R_p+R_t$,  $\overline{R}  = \dfrac{R_p R_t}{R_p+R_t}$ and $a$ is the diffuseness parameter $a= 0.63 fm$. This form is similar to that of the Bass model \cite{Bass} with different sets of radius parameter.
    \begin{equation}
    R_i = 1.233 A_i^\frac{1}{3}-0.98 A_i^\frac{1}{3} fm   (i=p,t)
     \label{eq20}
     \end{equation}
     Here, the radius of the fusion barrier has the form 
       \begin{equation}
       R_B = 1.07 (A_p^\frac{1}{3}+A_t^\frac{1}{3} )+2.72  fm
        \label{eq21}
        \end{equation}
        and the total nucleus-nucleus potential is 
        \begin{equation}
            U^{CW} (R) = \dfrac{Z_{p}Z_{t}e^{2}}{R} + V_N^{CW} (R)
             \label{eq22}
             \end{equation}
        and thus, the fusion barrier can be obtained from $U^{CW}(R=R_B$).
        
       \subsection{Broglia and Winther model}  
       Broglia and Winther \cite{Reisdorff} have refined the CW potential \cite{Christensen} in order to make it compatible with the value of the maximum nuclear force of the proximity potential \cite{Blocki}. This refined force is taken as the standard Woods-Saxon potential given by
        \begin{equation}
        V_N(R) = \dfrac{-V_{0}}{1+e^\frac{R-R_{pt}}{a}} MeV
         \label{eq23}
         \end{equation}
         with $V_0 = 16{\pi}a\gamma\frac{R_{p}R_{t}}{R_{p}+R_{t}} , a=0.63 fm$ and
         \begin{equation}
           R_{pt} = R_p+R_t+0.29 fm  
           \label{eq24}
         \end{equation}
         Here the nucleus radius $R_{i}$ is given by:
          \begin{equation}
          R_i = 1.233A_i^\frac{1}{3}-0.98A_i^\frac{-1}{3}  fm      (i=p,t)
          \label{eq25}
          \end{equation}
          The surface energy coefficient ($\gamma$) has the form
           \begin{equation}
           \gamma = \gamma_0 \bigg[1-k_s \bigg(\frac{N_{p}-Z_{p}}{A_{p}}\bigg)\bigg(\frac{N_{t}-Z_{t}}{A_{t}}\bigg)\bigg] MeV fm^2
            \label{eq26}
            \end{equation}
            Where $\gamma_0$ = $0.95 MeVfm^{-2}$  and $k_s = 1.8$.
            The total interaction potential of the two heavy ions is
            \begin{equation}
            U^{BW} (R) = \dfrac{Z_{p}Z_{t}e^{2}}{R} + V_N^{BW} (R)
             \label{eq27}
             \end{equation}
              and it displays a maximum, i.e., the fusion barrier and the barrier radius $(R_B )$ is the solution of the following equation
              \begin{equation}
             \dfrac{dU^{BW}(R)}{dR}|_{R=R_B} =-\dfrac{Z_{p}Z_{t}e^{2}}{R_{B}^{2}}+ \dfrac{V_{0}\hspace{2mm} e^\frac{R_{B}-R_{pt}}{a}}{{a\hspace{1mm} (1+e^{\frac{R_{B}-R_{pt}}{a}})}^{2}}=0
                        \label{eq28}
               \end{equation}
               and  $ U^{BW} (R_B )$ is the fusion barrier.
               
                \subsection{Aage Winther model}
               Aage Winther \cite{AW} adjusted slightly the parameters of the Broglia and Winther potential through an extensive comparison with the experimental data for heavy-ion elastic scattering. The resulting values of $a$ and $R_i$ are as follows:
               \begin{equation}
               a = \Bigg[\dfrac{1}{1.17(1+0.53(A_p^{-\frac{1}{3}}+A_t^{-\frac{1}{3}} )) }\Bigg]  \hspace {2mm}fm
               \label{eq29}
               \end{equation}
               and 
               \begin{equation}
               R_i = 1.20A_i^\frac{1}{3}-0.09 \hspace{2mm}fm           \hspace{9mm} (i=p,t)
                \label{eq30}
                \end{equation}
                and $R_{pt}$ of the BW model is written as $ R_{pt}$ = $R_p+R_t$ only.
                
                \subsection{Siwek-Wilczyńska and Wilczyński model}
                Siwek-Wilczyńska and Wilczyński \cite{Siwek} have used a large number of reactions to determine an effective nucleus-nucleus potential for reliable prediction of the fusion barriers for the systems that are studied. In this approach the nucleus-nucleus potential is taken also as the Woods-Saxon shape is given in equation (\ref{eq23}).
                  Where $R_{pt}$ = $ R_p+R_t$ and $R_i =  R_c A_i^\frac{1}{3}$ $(i=p,t)$, the radius parameter $R_c$ is constant, $a$ is the diffuseness parameter and $V_0$ is the depth of the potential. The $V_0$ is given by
                    \begin{equation}
                    V_0 = V_0^{'}+ S_{cn}
                     \label{eq31}
                     \end{equation}
                     Where $S_{cn}$ is the shell correction energy \cite{Scn} and 
                     \begin{center}
                     $V_0^{'} = (M_p+ M_t- M_{cn} ) c^2+ C_{cn}-C_p-C_t$\\
                     \end{center}
                     
                     \begin{equation}
                     = Q_{fu}+C_{cn}-C_p-C_t
                     \label{eq32}
                     \end{equation}
                     Here $Q_{fu}$ is the ground state $Q$ value for fusion and $C_i$ are the Coulomb energies \cite{W D Myers} as follow
                      \begin{center}
                     $C_{cn} - C_p - C_t = C_0$
                     \end{center}
                      \begin{equation}
                     C_0 = 0.7054\Bigg[\dfrac{(Z_p+Z_t )^2}{(A_p+A_t )^\frac{1}{3}} - \dfrac{Z_p^2}{A_p^\frac{1}{3}}-\dfrac{Z_t^2}{A_t^\frac{1}{3}}\Bigg] \hspace{4mm} MeV
                     \label{eq33}
                     \end{equation}
                     Therefore, the equation (\ref{eq31}) can be written as
                     \begin{equation}
                     V_0= Q_{fu}+C_0+ S_{cn}
                      \label{eq34}
                      \end{equation}
                      For determination of the fusion barrier, one considers the nucleus-nucleus potential in the region $R > R_{pt}$ as
                      \begin{equation}
                      V(R) = V_N (R)+ \dfrac{Z_p Z_t e^2}{4\pi\epsilon_0 R}
                       \label{eq35}
                       \end{equation}
                        For the region $R < R_p+R_t, e^{\big[\frac{R-R_{pt}}{a}\big] }\longrightarrow0$, the nucleus-nucleus potential takes the form
                        \begin{equation}
                        V(R) = C_0-V_0= -Q_{fu}-S_{cn}
                         \label{eq36}
                         \end{equation}
                         Equation (\ref{eq35}) gives thus the fusion barrier. It has only two free parameters $R_c$ and $a$ as $V_0$ is known from equation (\ref{eq34}). These parameters are obtained by fitting the barrier heights from equation (\ref{eq35}). The experimental $B_{fu}$ values can be obtained where the measured fusion excitation functions are filled with the following expression.
                           \begin{equation}
                           \sigma_{fus} = \pi r_{\sigma}^{2} \dfrac{\omega}{E\surd2\pi} [X\surd\pi (1+erf(X))+exp(-X^{2}]
                            \label{eq37}
                            \end{equation}
                            Where $X = \dfrac{E-B_{fu}}{\sqrt{2} \omega}$ and Gaussian error integral function of the argument  is $X$
                            \begin{equation}
                        erf(X) = \dfrac{1}{\sqrt{\pi}}{ \int_{0}^{X} e^{-t^{2}}dt}
                             \label{eq38}
                             \end{equation}
                             The fitting gives three parameters the fusion barrier $B_{fu}$, the relative distance corresponding to the position of the approximate barrier $r_\sigma$ and the width of the barrier $w$.
                             However, the values of $R_c$ and $a$ depend on the Coulomb barrier parameter $z$, for example,\\
                            $ R_c$ = $1.18 fm$ and $a = 0.675 fm$    \hspace{5mm}   for    $70<z<130$\\
$R_c =1.25 fm$ and $a=0.481 fm$\hspace{5mm} for$ z<70$\\
$R_c=1.11 fm$ and $a=0.895 fm$ \hspace{5mm}for $z>130$

\subsection{Skyrme Potential model}
Skyrme energy density function model (SEDFM) has been introduced by Wang \cite{Wang,Liu}. The total binding energy of a nucleus can be represented as the integral of the energy density function \cite{J Bartel} 
\begin{equation}
E = \int{Hdr}
\label{eq39}
\end{equation}
where energy-density function $H$ has three parts: kinetic energy, Coulomb and nuclear interactions and is generally defined as follows
\begin{equation}
H(r)=\frac{{\hbar}^2}{2m}\left[ \tau_{p}\left( r\right) +\tau_{n}\left( r\right) \right]+H_{coulomb}\left( r\right) +H_{nuclear}\left( r\right) 
\label{eq40}
\end{equation}
Where $\tau$ is the kinetic energy density.
The interaction potential $V_{B}\left( r\right) $ is defined as
\begin{equation}
V_{B}\left( R\right)  = E_{tot}\left( R\right) - E_{p} - E_{t}
\label{eq41}
\end{equation}
where $E_{tot}\left( R\right) $ is the total energy of the interacting nuclear system. $E_{p}$ and $E_{t}$ are the energies of the projectile and target at completely separated distance $R$, respectively. These energies can be calculated by the following relation
\begin{equation*}
 E_{tot}(R) =\int{H[ \rho_{1p}(r) +\rho_{2p}( r-R),\rho_{1n}(r) +\rho_{2n}( r-R)] dr}   
\end{equation*}
\begin{center}
$E_{p}\left( R\right) =\int{H\left[ \rho_{1p}\left( r\right) ,\rho_{1n}\left( r\right) \right] dr}$
\end{center}
\begin{equation}
E_{t}\left( R\right) =\int{H\left[ \rho_{2p}\left( r\right) ,\rho_{2n}\left( r\right) \right] dr}
\label{eq42}
\end{equation}
The densities of the neutron $\rho_{n}$ and proton $\rho_{p}$ for projectile and target can be described by the spherically symmetric Fermi function
\begin{equation}
\rho{\left( r\right)} =\dfrac{{\rho}_{0}}{1+{e^{\frac{\left( r-c\right)}{a}}}}
\label{eq43}
\end{equation}
where $\rho_{0}$, $c$ and $a$ are the parameters of the densities of the participating nuclei in the reactions, which are obtained by using the density-variation approach and minimizing the total energy of a single nucleus given by the SEDFM \cite{Wang, Liu}.

Using the Skyrme energy density formalism, Zanganeh et al. \cite{Zanganeh} have constructed a pocket formula for fusion barrier heights and positions in the range 8$\le$ z $\le$ 168 with respect to the charge and mass numbers of the interacting nuclei as follows:
\begin{equation}
    V_B^{Par}=-0.01[(Z_pZ_t)(A_p^{\frac{1}{3}}+A_t^{\frac{1}{3}})]+0.20(Z_p Z_t)+0.60
    \label{eq44}
\end{equation}
\begin{equation}
    R_B^{Par}=1.40[A_p^\frac{1}{3} + Z_t^\frac{1}{3}] - 0.07(Z_p Z_t)^{0.05} +1.40
    \label{eq45}
\end{equation}
We have made use of the equation (\ref{eq44},\ref{eq45}) for SEDFM predictions for various reactions as shown in Table 2 and Table 3. Since, the SEDFM is based on the frozen density approximation, predicted values for each of the considered fusion systems are a bit higher than the corresponding experimental data. 

\subsection{Sao Paulo optical potential} 
This model \cite{Freitas} also takes a Woods-Saxon form for the nuclear potential as given in equation (\ref{eq23}). In the approximation of $\exp{(\frac{R_B-R}{a})}\gg 1$, the $R_B$ can be written as:
\begin{equation}
R_B = R + 0.65 \ln{x}
\label{eq48}
\end{equation}
where $x = 27.1 \times \frac{A_p^\frac{1}{3}+A_t^\frac{1}{3}}{Z_p Z_t}$ is a positive dimensionless parameter. Note that the parameter $x$, which appears in the argument of the logarithm of the above equation, can also be written as $x = \exp({\frac{R_B-R}{a}})$ and is larger than one in most cases. The barrier potential $V_B$ is given by
\begin{equation}
    V_B = \frac{Z_p Z_t e^2}{R_B} - \frac{15}{x+1}
    \label{eq49}
\end{equation}
\begin{table*}[h!]
  \centering
  \caption{Comparison of the fusion barrier height $B_{fu}$ for different systems with experimental as well as various theoretical models. The reactions are listed in order of the increasing value of the $z$ parameter. The values for our model is taken from the equation (\ref{eq5}). }
  \begin{tabular}{p{2cm} p{1.5cm} p{2cm} p{1.5cm} p{1.5cm} p{1.5cm} p{1.5cm} p{1.5cm} p{1.5cm} p{1.5cm} p{1.0cm}}
    \hline
 \hline
    System&$z$&$Expt.$&$Skyrem$&$Bass$&$Poland$&$CW$&$BW$&$AW$&$SPP$&$Present$   \\
    \hline\\
    $^{12}$C+$^{14}$N &8.93 &7.00\hspace{2mm} \cite{Louis C. Vaz}&$7.03$&$5.19$& $7.14$&$7.06$&$7.11$&$7.04$&$6.73
$&$7.46$     \\
$^{12}$C+$^{18}$O & 9.77 & 7.45\hspace{2mm} \cite{Sperr}&$7.84$&$5.93$& $7.85$&$7.80$&$7.86$&$7.78$& $7.53
$&$8.18$\\
 $^{12}$C+$^{17}$O& $9.88$ & $8.20$ \hspace{0.5mm} \cite{V. Zanganeh}&$7.97$&$5.97$& $7.86$&$7.89$&$7.94$&$7.86$& $7.59$&$8.27$   \\ 
 $^{20}$Ne+$^{20}$Ne& $18.42$ & $15.20$ \cite{D. Shapira}&$16.02$&$13.15$& $15.57$&$15.59$&$15.69$&$15.61$& $15.46
$& $15.77$   \\
  $^{4}$He+$^{164}$Dy& $18.69$ & $17.14$ \cite{V. Zanganeh}&$17.09$&$15.19$& $17.67$&$17.19$&$17.30$&$17.09$& $16.44$ &$16.01$   \\
 $^{18}$O+$^{28}$Si& $19.79$ &$16.90$\hspace{1mm}\cite{V. Zanganeh} &$17.41$&$14.51$&$16.89$&$16.89$&$17.01$&$16.94$&$16.85$&$16.99$   \\
 $^{16}$O+$^{28}$Si&$20.16$ &$17.23$ \cite{V. Zanganeh}&$17.75$&$14.76$&$17.34$&$17.23$&$17.34$&$17.26$&$17.14
$ &$17.31$   \\
 $^{24}$Mg+$^{24}$Mg&$24.96$ &$22.30$ \cite{C. M. Jachcinski}&$22.47$&$19.16$& $21.63$&$21.69$&$21.82$&$21.76$&$21.79$&$21.59$   \\
 $^{6}$Li+$^{144}$Sm&$26.35$ &$24.65$ \cite{V. Zanganeh}&$24.65$&$22.22$& $24.19$&$24.32$&$24.47$&$24.33$&$23.78
$&$22.84$   \\
 $^{7}$Li+$^{159}$Tb&$26.60$ &$23.81$ \cite{V. Zanganeh}&$25.00$&$22.58$&$24.44$&$24.45$&$24.73$&$24.64$&$24.14
$ &$23.07$   \\
  $^{16}$O+$^{40}$Ca&$26.94$ &$23.70$ \cite{C. M. Jachcinski}&$24.52$&$21.18$&$23.69$&$23.68$&$23.81$&$23.75$&$23.78
$& $23.37$   \\
 $^{26}$Mg+$^{30}$Si&$27.68$ &$24.80$ \cite{V. Zanganeh}&$25.29$&$21.88$&$24.25$&$24.29$&$24.46$&$24.42$&$24.57
$& $24.04$   \\
 $^{14}$N+$^{59}$Co&$29.99$ &$26.13$ \cite{PRS}&$27.75$&$24.45$&$26.68$&$26.79$&$26.95$&$26.90$&$26.95
$&$26.12$   \\
 $^{26}$Mg+$^{34}$S&$30.96$ &$27.11$ \cite{G. M. Berkowitz}&$28.62$&$25.06$&$27.64$&$27.42$&$27.61$&$27.59$&$24.07
$&$27.00$   \\
$^{24}$Mg+$^{34}$S&$31.35$ &$27.38$ \cite{G. M. Berkowitz}&$28.95$&$25.39$&$27.69$&$27.80$&$27.96$&$27.92$&$24.37
$&$27.36$   \\
 $^{30}$Si+$^{30}$Si&$31.53$ &$28.54$ \cite{E. F. Aguilera}&$29.19$&$25.62$&$27.93$&$27.97$&$28.16$&$28.14$&$28.42
$&$27.53$   \\
 $^{24}$Mg+$^{32}$S&$31.69$ &$28.10$ \cite{G. M. Berkowitz}&$29.23$&$25.65$& $28.01$&$28.09$&$28.25$&$28.21$&$24.61
$&$27.66$   \\
 $^{6}$Li+$^{208}$Pb&$31.77$ &$30.10$\hspace{1mm}\cite{V. Zanganeh} &$30.27$&$28.20$&$30.12$&$29.97$&$30.18$&$30.06$&$29.40
$& $27.74$   \\
 $^{28}$Si+$^{30}$Si&$31.90$ &$29.13$ \cite{S. Gary}&$29.49$&$25.91$&$28.25$&$28.31$&$28.47$&$28.44$&$28.74
$&$27.86$   \\
$^{28}$Si+$^{28}$Si&$32.27$ &$28.89$ \cite{S. Gary}&$29.85$&$26.22$&$28.63$&$28.64$&$28.80$&$28.76$&$29.06
$&$28.19$   \\
 $^{20}$Ne+$^{40}$Ca&$32.60$ &$28.60$ \cite{V. Zanganeh}&$30.22$&$26.64$&$28.88$&$29.04$&$29.20$&$29.16$&$29.42
$&$28.50$   \\
 $^{24}$Mg+$^{35}$Cl&$33.14$ &$30.70$ \cite{V. Zanganeh}&$30.76$&$27.13$&$29.37$&$29.51$&$29.68$&$29.64$&$29.96
$&$28.98$   \\
 $^{16}$O+$^{58}$Ni&$35.05$ &$31.67$ \cite{V. Zanganeh}&$32.87$&$29.36$&$31.64$&$31.61$&$31.79$&$31.75$&$31.99
$&$30.73$   \\
 $^{12}$C+$^{152}$Sm&$48.78$ &$46.39$ \cite{V. Zanganeh}&$47.72$&$44.97$&$45.75$&$46.06$&$46.38$&$46.38$&$46.50
$&$43.42$   \\
$^{18}$O+$^{124}$Sn&$52.58$ &$49.30$ \cite{V. Zanganeh}&$51.57$&$48.48$&$48.38$&$49.31$&$49.82$&$49.97$&$50.40
$&$46.98$   \\
$^{16}$O+$^{116}$Sn&$54.08$ &$50.94$ \cite{V. Zanganeh}&$52.80$&$50.00$&$50.51$&$50.87$&$51.20$&$51.29$&$51.84
$&$48.39$   \\
$^{30}$Si+$^{64}$Ni&$55.15$ &$51.20$ \cite{A.M. sTefanini}&$53.90$&$50.25$&$50.85$&$51.20$&$51.59$&$51.71$&$52.82
$&$49.41$   \\
 $^{30}$Si+$^{62}$Ni&$55.48$ &$52.20$ \cite{A.M. sTefanini}&$54.24$&$50.56$&$51.19$&$51.51$&$51.87$&$51.99$&$53.14
$&$49.72$   \\
 $^{28}$Si+$^{64}$Ni&$55.71$ &$52.40$ \cite{A.M. sTefanini}&$54.35$&$50.82$&$51.29$&$51.76$&$52.08$&$52.18$&$53.36
$&$49.93$   \\
$^{28}$Si+$^{62}$Ni&$56.04$ &$52.89$ \cite{A.M. sTefanini}&$54.74$&$51.14$&$51.67$&$52.07$&$52.38$&$52.47$&$53.68
$&$50.25$   \\
$^{40}$Ca+$^{48}$Ca&$56.70$ &$52.00$ \cite{V. Zanganeh}&$55.21$&$51.73$&$51.97$&$52.61$&$52.95$&$53.07$&$54.40
$&$50.87$   \\
$^{28}$Si+$^{58}$Ni&$56.75$ &$53.80$\hspace{1mm}\cite{A.M. sTefanini}&$55.46$&$51.82$&$52.39$&$52.71$&$53.00$&$53.08$&$54.37
$&$50.92$   \\
 $^{12}$C+$^{204}$Pb&$60.17$ &$57.55$ \cite{V. Zanganeh}&$59.84$&$57.89$&$57.88$&$57.92$&$58.33$&$58.50$&$58.53
$&$54.17$   \\
$^{40}$Ca+$^{48}$Ti&$62.37$ &$58.17$ \cite{A. A. Sonzogni}&$61.38$&$57.82$&$57.56$&$58.22$&$58.56$&$58.68$&$60.32
$&$56.26$   \\
$^{35}$Cl+$^{54}$Fe&$62.69$ &$58.59$ \cite{V. Zanganeh}&$61.89$&$58.20$&$58.08$&$58.56$&$58.90$&$59.02$&$60.65
$&$56.56$   \\
$^{16}$O+$^{144}$Sm&$63.91$ &$61.03$ \cite{JR Leigh}&$63.62$&$61.08$&$60.60$&$61.00$&$61.39$&$61.56$&$62.25
$&$57.73$   \\
$^{37}$Cl+$^{64}$Ni&$64.92$ &$60.60$ \cite{V. Zanganeh} &$64.32$&$60.80$&$60.32$&$60.90$&$61.36$&$61.56$&$63.12
$&$58.70$   \\
 $^{46}$Ti+$^{46}$Ti&$67.54$ &$63.30$ \cite{V. Zanganeh}&$66.99$&$63.50$&$62.66$&$63.39$&$63.76$&$63.92$&$65.81
$&$61.21$   \\
 $^{16}$O+$^{186}$W&$71.95$ &$68.87$ \cite{JR Leigh}&$72.26$&$70.53$&$68.66$&$69.48$&$69.99$&$70.27$&$70.91
$&$65.46$   \\
    $^{28}$Si+$^{92}$Zr&$74.16$ &$70.93$ \cite{JR Newton}&$74.24$&$71.43$&$69.17$&$70.50$&$70.93$&$71.16$&$72.99
$&$67.60$   \\
$^{40}$Ca+$^{58}$Ni&$76.81$ &$73.36$ \cite{L. C. Vaz}&$76.96$&$73.90$&$71.54$&$72.71$&$73.11$&$73.30$&$75.70
$&$70.17$   \\
$^{16}$O+$^{208}$Pb&$77.68$ &$74.90$ \cite{C. R. Morton}&$78.46$&$77.25$&$75.39$&$75.50$&$76.07$&$76.42$&$77.07
$&$71.02$   \\
$^{48}$Ti+$^{58}$Ni&$82.08$ &$78.80$ \cite{A. M. Vinodkumar}&$82.74$&$79.86$&$76.76$&$78.09$&$78.57$&$78.83$&$81.42
$&$75.31$   \\
$^{36}$S+$^{90}$Zr&$82.23$ &$79.00$ \cite{A. M. Stefanini}&$83.17$&$80.39$&$77.48$&$78.55$&$79.18$&$79.51$&$81.68
$&$75.46$   \\
$^{19}$F+$^{197}$Au&$83.77$ &$81.61$ \cite{V. Zanganeh}&$85.20$&$83.88$&$80.73$&$81.47$&$82.18$&$82.61$&$83.58
$&$76.98$   \\
 $^{35}$Cl+$^{92}$Zr&$87.34$ &$82.94$ \cite{JR Newton}&$88.60$&$86.25$&$82.31$&$83.76$&$84.32$&$84.66$&$87.17
$&$80.48$   \\
 $^{35}$Cl+$^{106}$Pd&$97.70$ &$94.30$ \cite{V. Zanganeh}&$100.02$&$98.36$&$92.86$&$94.48$&$95.11$&$95.54$&$98.43
$&$90.74$   \\
$^{32}$S+$^{116}$Sn&$99.36$ &$97.36$ \cite{V. Zanganeh}&$101.50$&$100.47$&$94.79$&$96.31$&$96.92$&$97.35$&$100.24
$&$92.39$   \\
$^{58}$Ni+$^{60}$Ni&$100.69$ &$96.00$ \cite{V. Zanganeh}&$103.17$&$101.33$&$95.82$&$97.04$&$97.63$&$98.01$&$101.64
$&$93.73$   \\
$^{40}$Ca+$^{90}$Zr&$101.25$ &$96.88$ \cite{V. Zanganeh}&$103.75$&$102.26$&$96.34$&$97.87$&$98.45$&$98.86$&$102.26
$&$94.28$   \\
$^{58}$Ni+$^{58}$Ni &101.27 &95.8\hspace{2mm}  \cite{Timmers}&$96.70$&$102.01$&$96.50$&$97.59$&$98.15$&$98.51$&$102.25
$&$94.30$ \\
$^{40}$Ar+$^{122}$Sn &107.40 &103.6  \cite{Beckerman}&$105.18$&$109.70$&$103.24$&$104.51$&$105.47$&$106.09$&$109.07
$&$100.44$\\
$^{40}$Ar+$^{116}$Sn &108.47 &103.3  \hspace{0mm}\cite{Reisdorf}&$105.93$&$110.92$&$104.26$&$105.57$&$106.45$&$107.02$&$110.22
$&$101.52$\\
$^{40}$Ar+$^{112}$Sn &109.22 &104.0 \cite{Reisdorf}&$106.44$&$111.78$&$105.02$&$106.30$&$107.12$&$107.67$&$111.02
$&$102.27$\\
$^{64}$Ni+$^{74}$Ge &109.29 &103.2 \cite{Reisdorf}&$106.34$&$111.37$&$104.46$&$106.00$&$106.92$&$107.50$&$111.09
$&$102.34$\\
$^{40}$Ar+$^{121}$Sb & 109.72 & 111.0 \cite{beckerman}&$107.40$&$112.44$&$105.59$&$106.90$&$107.84$&$108.45$&$111.60
$&$102.78$\\
$^{58}$Ni+$^{74}$Ge & 111.04 & 106.8 \cite{beckerman}&$107.50$&$113.49$&$105.97$&$107.76$&$108.49$&$109.01$&$112.99
$&$104.10$\\ 
$^{34}$S+$^{168}$Er & 124.24 &121.5  \cite{Hagino}&$122.92$&$130.30$&$121.20$&$122.46$&$123.51$&$124.30$&$127.56
$&$117.46$\\
$^{28}$Si+$^{208}$Pb&$128.10$&$128.07$ \hspace{-1.5mm}\cite{D. J. Hinde}&$133.48$&$135.65$&$126.22$&$127.07$&$128.03$&$128.87$&$131.83
$&$121.40$   \\
$^{40}$Ar+$^{148}$Sm &128.14 &124.7 \cite{Reisdorf}&$126.60$&$134.57$&$124.99$&$126.19$&$127.32$&$128.13$&$131.85
$&$121.43$\\
$^{40}$Ar+$^{144}$Sm &128.85 &124.4 \cite{Reisdorf}&$127.14$&$135.41$&$125.91$&$126.90$&$127.98$&$128.76$&$132.63
$&$122.16$\\
    
  \end{tabular}
  \end{table*}
   \begin{figure*}[h!]
\centering
\includegraphics[width=\linewidth,]{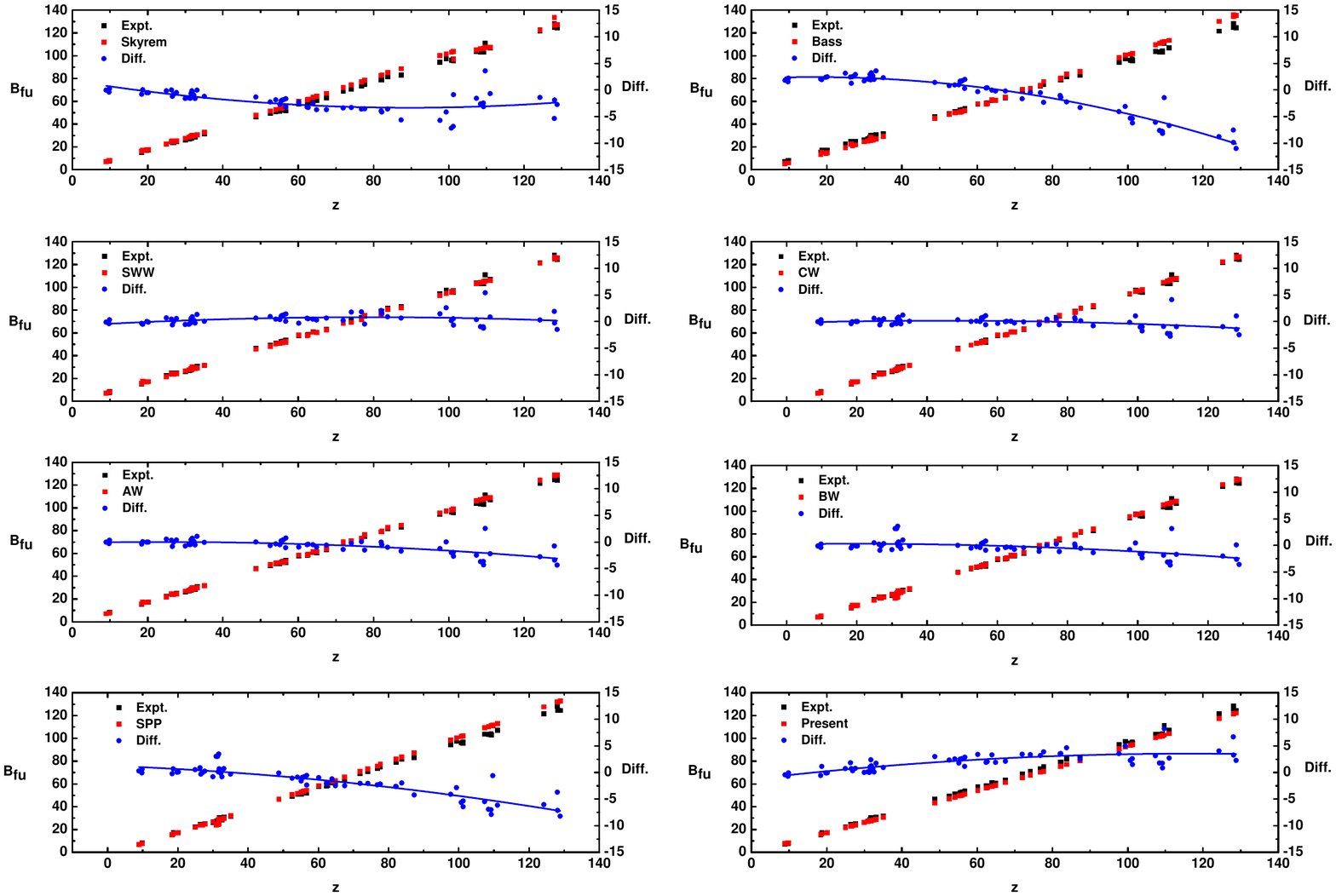} 
\caption{Comparison of experimental fusion barrier heights with the present empirical formula and different theoretical models as a function of $z$. Difference between the experiment and every theoretical model are also plotted and fitted with a second order polynomial function. }
\end{figure*}
\begin{table*}
  \centering
  \caption{Comparison of the fusion barrier radius $R_{B}$  for different systems with experimental as well as various theoretical models. The reactions are listed in order of the increasing value of the $z$ parameter. The values for our model is taken from the equation (\ref{eq11}). The superscript * indicates that the experimental radii for all the reactions are taken from the ones given in \cite{Siwek}.}
  \begin{tabular}{p{2cm} p{1.5cm} p{2cm} p{1.5cm} p{1.5cm} p{1.5cm} p{1.5cm} p{1.5cm} p{1.5cm} p{1.5cm} p{1.1cm}}
    \hline
    \hline
System&$z$&$Expt.^*$&$Skyrem$&$Bass$&$Poland$&$CW$&$BW$&$AW$&$SPP$&$Present$   \\
    \hline \\
    $^{48}$Ca+$^{48}$Ca &55.03 &11.20&$10.18$&$10.10$& $11.30$&$10.50$&$10.46$&$10.37$&$10.28$&$11.09$     \\
$^{30}$Si+$^{64}$Ni & 55.16 & 9.60&$9.96$&$9.82$& $10.60$&$10.32$&$10.21$&$10.16$& $10.05
$&$10.84$\\
 $^{30}$Si+$^{62}$Ni& $55.48$ & $9.70$&$9.91$&$9.75$& $10.50$&$10.28$&$10.15$&$10.10$& $9.99$&$10.77$   \\ 
 $^{28}$Si+$^{64}$Ni& $55.71$ & $7.60$ &$9.87$&$9.69$& $10.50$&$10.25$&$10.11$&$10.03$& $9.95
$& $10.72$   \\
  $^{28}$Si+$^{62}$Ni& $56.04$ & $7.70$ &$9.81$&$9.62$& $10.40$&$10.20$&$10.03$&$9.97$& $9.88$ &$10.65$   \\
 $^{30}$Si+$^{58}$Ni& $56.18$ &$8.80$ &$9.78$&$9.60$&$10.40$&$10.19$&$10.02$&$9.94$&$9.86$&$10.62$   \\
 $^{40}$Ca+$^{48}$Ca&$56.70$ &$11.50$&$9.88$&$9.72$&$10.50$&$10.27$&$10.15$&$10.09$&$9.96
$ &$10.73$   \\
 $^{28}$Si+$^{58}$Ni&$56.75$ &$8.10$ &$9.68$&$9.47$& $10.20$&$10.11$&$9.89$&$9.84$&$9.76$&$10.51$   \\
 $^{40}$Ca+$^{44}$Ca&$57.55$ &$7.90$ &$9.73$&$9.53$& $10.30$&$10.16$&$9.98$&$9.90$&$9.81
$&$10.56$   \\
 $^{40}$Ca+$^{40}$Ca&$58.48$ &$9.50$ &$9.58$&$9.33$&$10.20$&$10.04$&$9.79$&$9.73$&$9.64
$ &$10.38$   \\
  $^{36}$S+$^{64}$Ni&$61.35$ &$8.50$ &$10.14$&$9.90$&$10.90$&$10.53$&$10.43$&$10.34$&$10.25
$& $11.03$   \\
 $^{34}$S+$^{64}$Ni&$61.88$ &$8.50$ &$10.05$&$9.79$&$10.70$&$10.47$&$10.33$&$10.25$&$10.16
$& $10.93$   \\
 $^{40}$Ca+$^{50}$Ti&$61.94$ &$9.40$ &$9.88$&$9.61$&$10.60$&$10.32$&$10.15$&$10.07$&$9.97
$&$10.73$   \\
 $^{40}$Ca+$^{48}$Ti&$62.37$ &$9.40$&$9.81$&$9.52$&$10.40$&$10.27$&$10.07$&$10.00$&$9.90
$&$10.64$   \\
$^{32}$S+$^{64}$Ni&$62.44$ &$8.10$ &$9.96$&$9.67$&$10.70$&$10.40$&$10.23$&$10.16$&$10.07
$&$10.83$   \\
 $^{36}$Si+$^{58}$Ni&$62.46$ &$7.70$&$9.96$&$9.68$&$10.60$&$10.39$&$10.23$&$10.13$&$10.06
$&$10.82$   \\
 $^{40}$Ca+$^{46}$Ti&$62.83$ &$9.40$&$9.74$&$9.42$& $10.40$&$10.21$&$9.98$&$9.92$&$9.82
$&$10.56$   \\
 $^{16}$O+$^{154}$Sm&$62.94$ &$9.60$ &$10.87$&$10.44$&$11.50$&$11.15$&$11.07$&$11.01$&$11.04
$& $11.88$   \\
 $^{34}$S+$^{58}$Ni&$63.01$ &$7.60$ &$9.87$&$9.57$&$10.50$&$10.33$&$10.13$&$10.06$&$9.97
$&$10.72$   \\
$^{17}$O+$^{144}$Sm&$63.49$ &$10.80$&$10.78$&$10.35$&$11.40$&$11.08$&$10.98$&$10.91$&$10.94
$&$11.77$   \\
 $^{16}$O+$^{148}$Sm&$63.51$ &$10.20$ &$10.77$&$10.33$&$11.40$&$11.08$&$10.97$&$10.90$&$10.93
$&$11.77$   \\
 $^{32}$S+$^{58}$Ni&$63.59$ &$8.30$ &$9.78$&$9.44$&$10.40$&$10.26$&$10.02$&$9.95$&$9.87
$&$10.61$   \\
 $^{16}$O+$^{144}$Sm&$63.91$ &$10.30$ &$10.71$&$10.25$&$11.30$&$11.02$&$10.89$&$10.82$&$10.86
$&$11.69$   \\
 $^{16}$O+$^{186}$W&$71.95$ &$10.60$ &$11.22$&$10.63$&$11.70$&$11.52$&$8.73$&$8.50$&$11.43
$&$12.26$   \\
$^{16}$O+$^{208}$Pb&$77.68$ &$10.50$ &$11.43$&$10.77$&$11.80$&$11.76$&$9.03$&$8.78$&$11.68
$&$12.49$   \\
$^{36}$S+$^{96}$Zr&$81.21$ &$11.00$ &$10.66$&$10.16$&$11.30$&$11.15$&$8.40$&$10.91$&$10.87
$&$11.61$   \\
$^{36}$S+$^{90}$Zr&$82.23$ &$10.80$ &$10.53$&$10.00$&$11.20$&$11.05$&$10.89$&$10.77$&$10.73
$&$11.45$   \\
 $^{36}$S+$^{110}$Pd&$90.94$ &$8.20$ &$10.83$&$10.23$&$11.50$&$11.38$&$8.72$&$8.45$&$11.09
$&$11.79$   \\
 $^{32}$S+$^{110}$Pd&$92.39$ &$8.00$ &$10.65$&$10.00$&$11.40$&$11.24$&$8.57$&$10.92$&$10.91
$&$11.59$   \\
$^{64}$Ni+$^{64}$Ni&$98.00$ &$7.80$ &$10.64$&$10.01$&$11.50$&$11.28$&$8.65$&$10.94$&$10.92
$&$11.57$   \\
$^{58}$Ni+$^{64}$Ni&$99.61$ &$6.5$ &$10.46$&9.79&$11.20$&$11.14$&$8.51$&$10.76$&$10.73
$&$11.37$   \\
$^{40}$Ca+$^{96}$Zr&$100.01$ &$9.30$&$10.62$&$9.93$&$11.40$&$11.28$&$8.65$&$10.91$&$10.90$&$11.55$   \\
 $^{58}$Ni+$^{60}$Ni&$100.70$ &$7.50$&$10.34$&$9.64$&$11.00$&$11.05$&$10.77$&$10.62$&$10.60
$&$11.23$   \\
$^{40}$Ca+$^{90}$Zr&$101.25$ &$10.00$ &$10.48$&$9.77$&$11.20$&$11.17$&$10.92$&$10.76$&$10.76
$&$11.3$   \\
$^{58}$Ni+$^{58}$Ni&$101.27$ &$6.00$&$10.28$&$9.56$&$10.90$&$11.00$&$10.70$&$10.55$&$10.54
$&$11.16$   \\
$^{40}$Ar+$^{122}$Sn&$107.40$ &$9.80$ &$11.03$&$10.33$&$11.80$&$11.69$&$9.15$&$8.87$&$11.38
$&$12.02$   \\
$^{40}$Ar+$^{116}$Sn&$108.47$ &$8.70$  &$10.92$&$10.19$&$11.70$&$11.60$&$9.06$&$8.79$&$11.26
$&$11.89$   \\
 $^{40}$Ar+$^{112}$Sn&$109.22$ &$8.90$ &$10.84$&$10.10$&$11.60$&$11.54$&$8.99$&$8.73$&$11.18
$&$11.80$   \\
 $^{64}$Ni+$^{74}$Ge&$109.29$ &$6.50$ &$10.78$&$10.09$&$11.60$&$11.49$&$8.95$&$8.67$&$11.12
$&$11.74$   \\
    $^{58}$Ni+$^{74}$Ge&$111.04$ &$7.00$ &$10.60$&$9.87$&$11.40$&$11.35$&$8.80$&$10.93$&$10.93
$&$11.53$   \\
$^{40}$Ca+$^{124}$Sn&$118.95$ &$9.60$ &$10.96$&$10.17$&$11.80$&$11.72$&$9.23$&$8.95$&$11.35
$&$11.94$   \\
$^{28}$Si+$^{198}$Pt&$123.18$ &$9.80$ &$11.50$&$10.64$&$12.30$&$12.21$&$9.79$&$9.49$&$11.96
$&$12.56$   \\
$^{34}$S+$^{168}$Er&$124.24$ &$10.30$ &$11.35$&$10.53$&$12.20$&$12.09$&$9.68$&$9.38$&$11.81$&$12.39$   \\
$^{40}$Ar+$^{154}$Sm&$127.11$ &$7.30$ &$11.35$&$10.56$&$12.20$&$12.11$&$9.72$&$9.42$&$11.82
$&$12.40$   \\
$^{40}$Ar+$^{148}$Sm&$128.14$ &$8.30$ &$11.25$&$10.44$&$12.10$&$12.04$&$9.65$&$9.34$&$11.72
$&$12.29$   \\
 $^{40}$Ar+$^{144}$Sm&$128.85$ &$8.30$ &$11.19$&$10.37$&$12.00$&$11.99$&$9.60$&$9.30$&$11.65
$&$12.22$   \\
 $^{40}$Ca+$^{192}$Os&$165.42$ &$10.70$ &$11.54$&$10.62$&$12.20$&$12.55$&$10.35$&$10.04$&$12.21
$&$12.76$   \\
$^{40}$Ca+$^{194}$Pt&$169.40$ &$9.60$ &$11.53$&$10.60$&$-$&$12.97$&$10.40$&$10.09$&$12.22
$&$12.77$   \\
\hline
\hline
  \end{tabular}
  \end{table*}
\begin{figure*}[h!]
\centering
\includegraphics[width=\linewidth,]{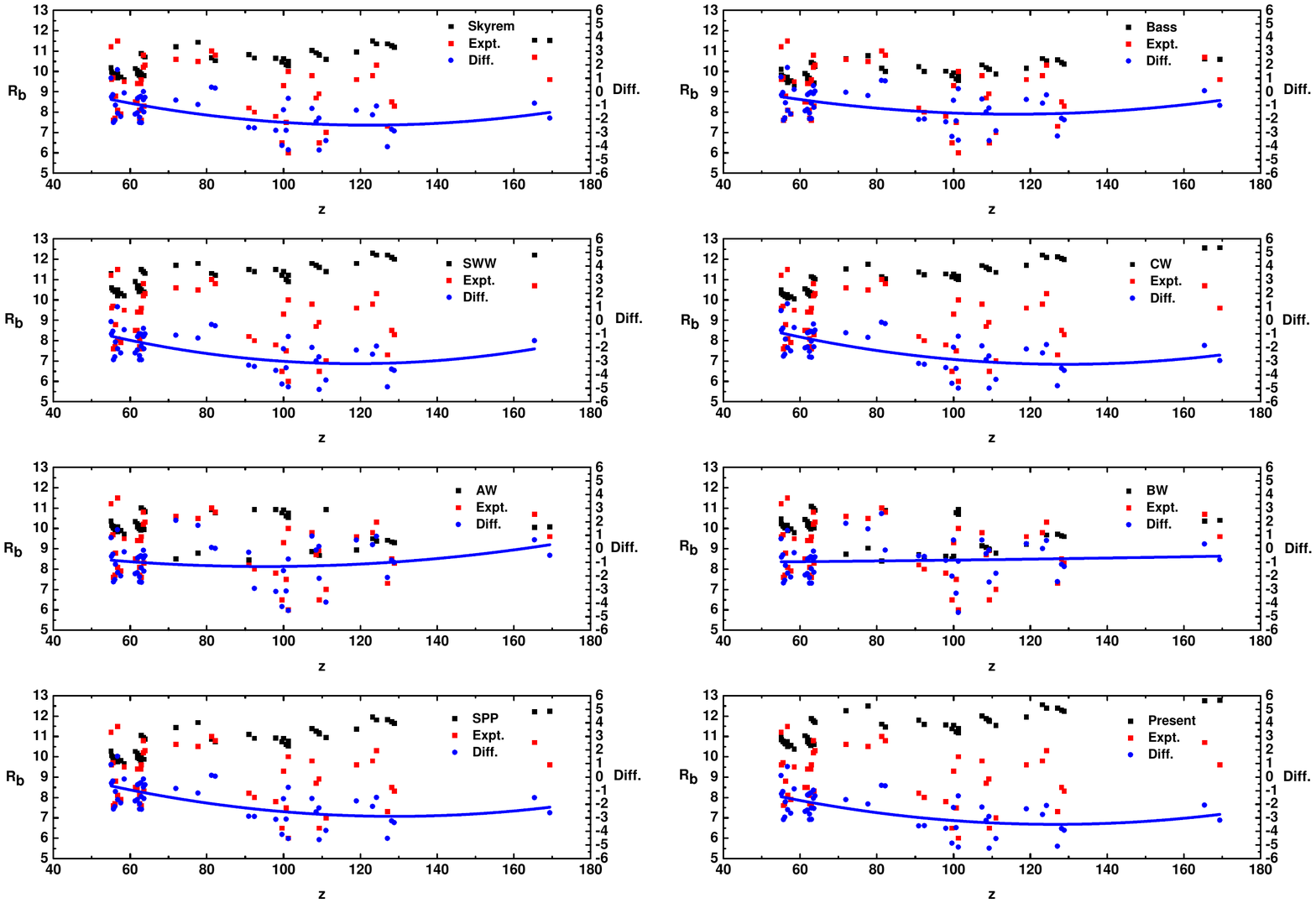}
\caption{Comparison of experimental fusion barrier radius with the present empirical formula and different theoretical models as a function of $z$. Difference between the experiment and every theoretical model are also plotted and fitted with a second order polynomial function.}
\end{figure*}

  \begin{figure}[h!]
\centering
\includegraphics[width=\linewidth,]{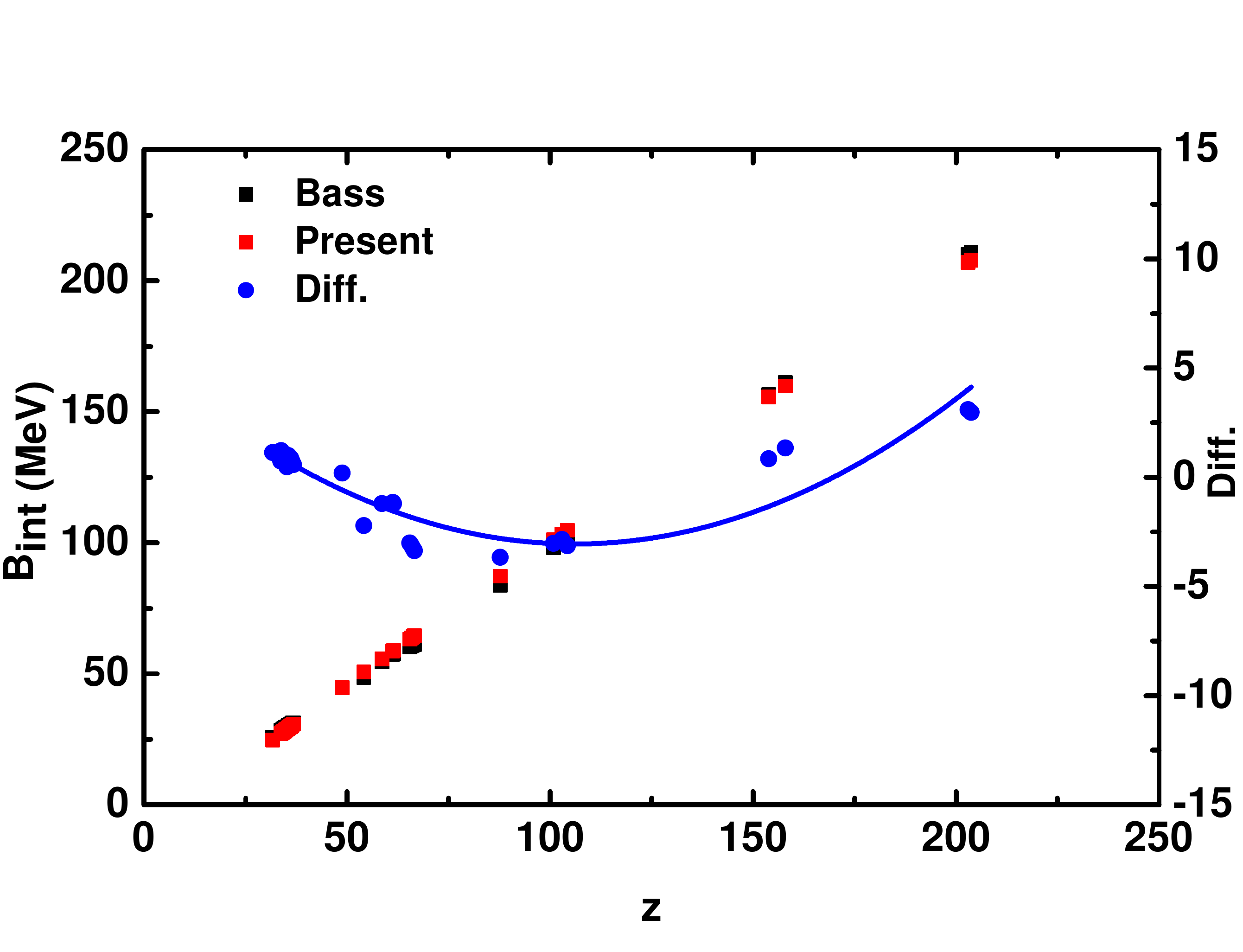} 
\caption{Comparison of interaction barrier heights between the present empirical formula and Bass model as a function of $z$. The difference between the predictions of the Bass model and present empirical model are also plotted and fitted with a second order polynomial function.}
\end{figure}.
\begin{figure*}[h!]
\centering
\includegraphics[width=\linewidth,]{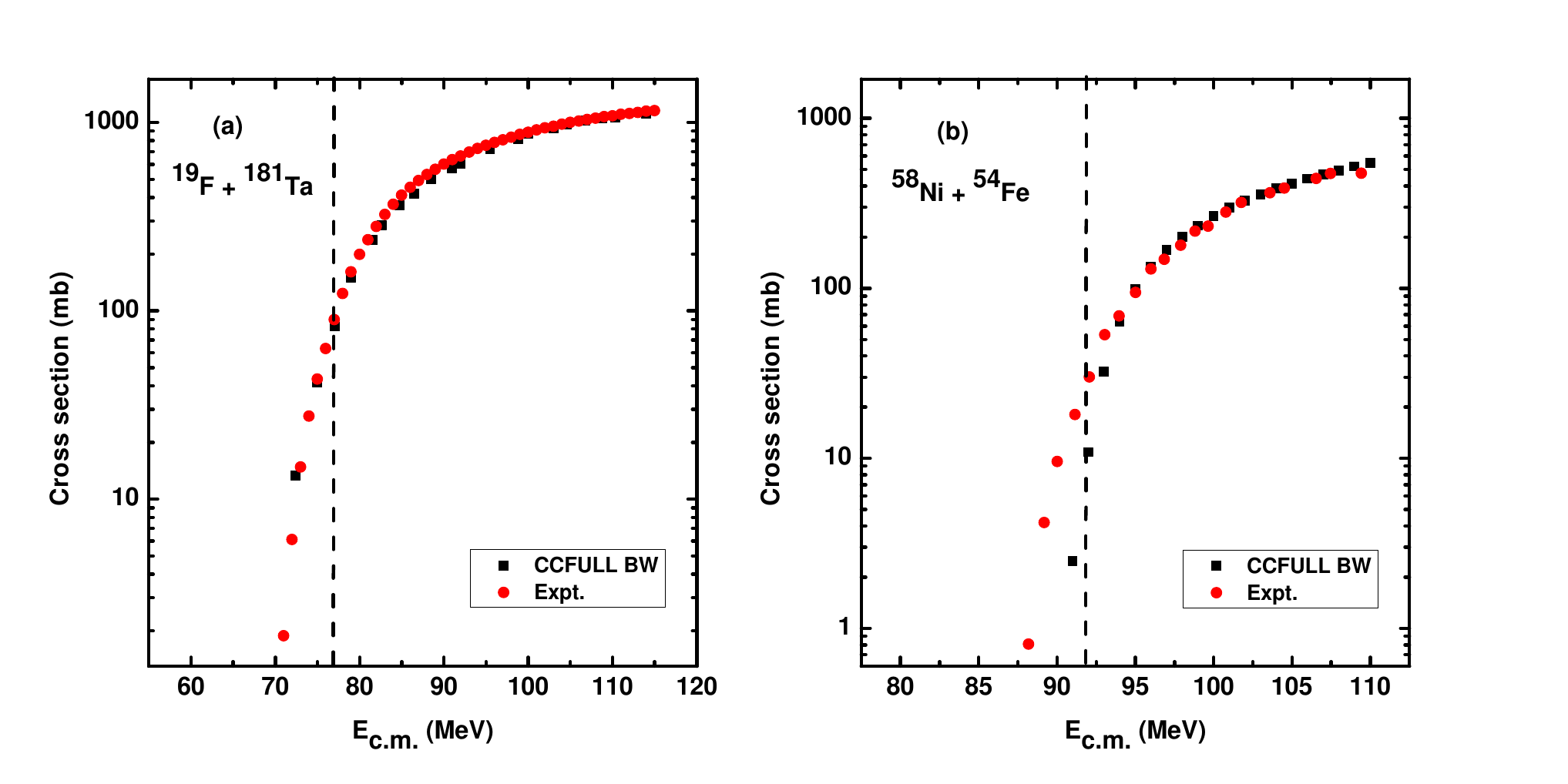} 
\caption{Comparison of total fusion cross section as a function of $E_{cm}$ between the experimental and CCFULL calculation using Broglia and Winther parameters for the systems of $^{19}$F+$^{181}$Ta (a) and $^{58}$Ni+$^{54}$Fe (b). The dashed vertical line indicates the fusion barrier height for the corresponding reaction.}
\end{figure*}
\begin{figure*}[h!]
\centering
\includegraphics[width=\linewidth,]{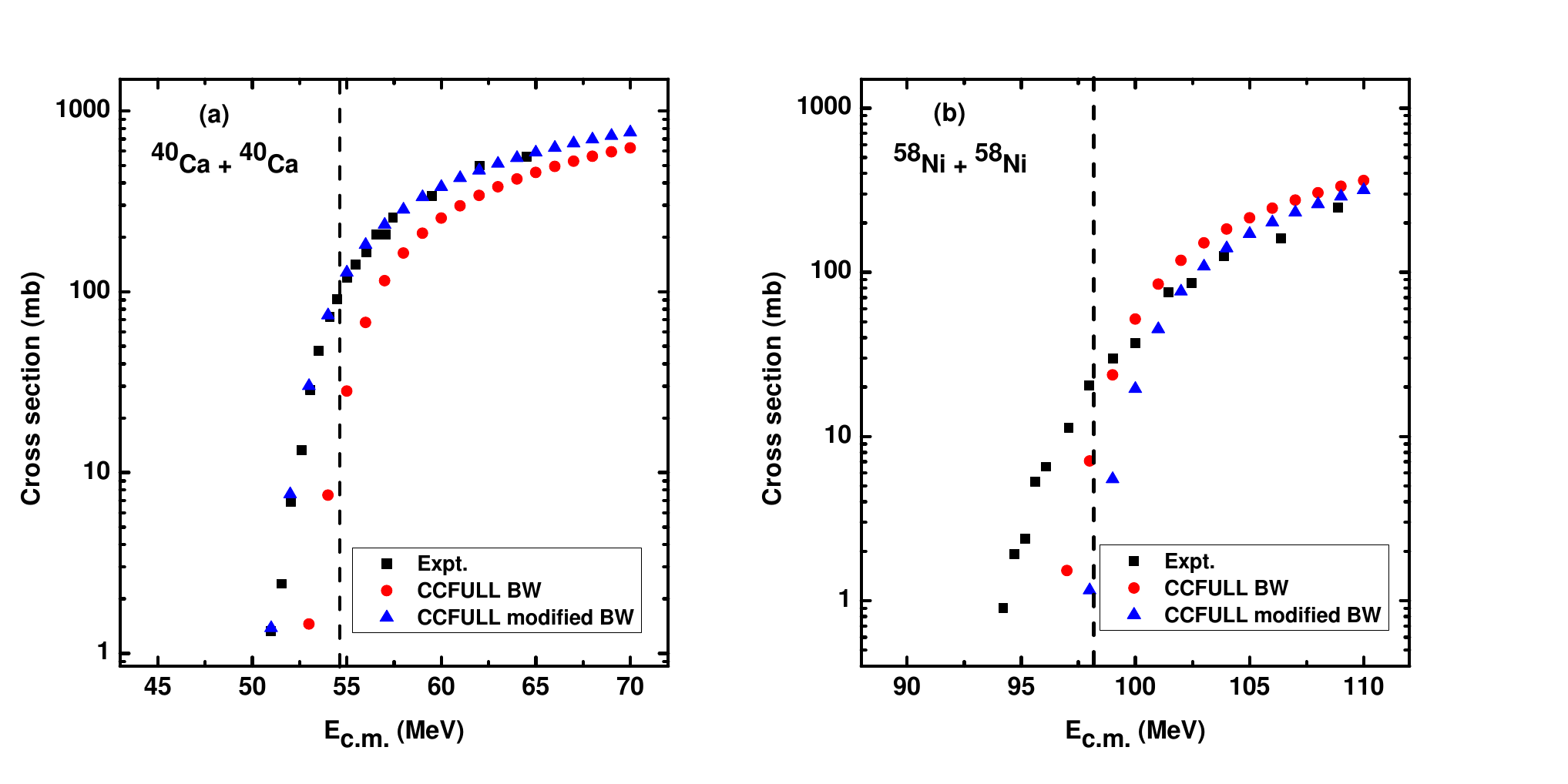} 
\caption{Comparison of total fusion cross section (mb) as a function of $E_{cm}$ between the experimental and CCFULL calculation using Broglia and Winther parameters for the systems of $^{40}$Ca+$^{40}$Ca (a) and $^{58}$Ni+$^{58}$Ni (b). Notice that the Broglia and Winther parameter $r_0$ does not result in agreement with the experimental values. Good concurrence is found by changing  $r_0$ from 1.19 to 1.24 fm for $^{40}$Ca+$^{40}$Ca and from 1.205 to 1.19 fm for $^{58}$Ni+$^{58}$Ni. The dashed vertical line indicates the fusion barrier height for the corresponding reaction.}
\end{figure*}
 \begin{table}[h!]
\caption{Comparison of the Woods-Saxon potential parameters $V_0$ (MeV), $r_0$ (fm), and $a_0$ (fm) between the BW model and others. Ichikawa et al. \cite{Ichikawa}, have quoted the potential energy at the ground state in place of the parameter $V_0$. }
\begin{center}
\centering
\begin{tabular}{|c|c|c|c|c|c|c|c|}
\hline
\multirow{2}{*}{Systems}  & \multicolumn{3}{c|}{B.W. Model} &\multicolumn{3}{c|}{Others} & \multirow{2}{*}{Ref.} \\
\cline{2-7}
&$V_0$&$a_0$&$r_0$&$V_0$&$a_0$&$r_0$&\\
\hline
$^{19}$F+$^{181}$Ta & 60.34 & 0.63 & 1.20 & 104.5 & 0.70 & 1.12 & \cite{Moin} \\
$^{48}$Ca+$^{96}$Zr & 67.67 & 0.63 & 1.209 & 104.5 & 0.68 & 1.198 & \cite{Ichikawa} \\
$^{16}$O+$^{154}$Sm & 57.45 & 0.63 & 1.197 & 100.0 & 1.06 & 1.019 & \cite{PRC2004}\\
$^{19}$F+$^{208}$Pb & 61.12 & 0.63 & 1.204 & 100.0 & 1.062 & 1.059 & \cite{PRC2004}\\
$^{16}$O+$^{208}$Pb & 59.20 & 0.63 & 1.201 (1.23) & 104.5 & 0.68 & 1.20 & \cite{Ichikawa} \\
$^{48}$Ca+$^{48}$Ca & 60.15 & 0.63 & 1.198 & 104.5 & 0.68 & 1.185 & \cite{Ichikawa} \\
$^{58}$Ni+$^{54}$Fe & 66.90 & 0.63 & 1.203 & 104.5 & 0.68 & 1.198 & \cite{Ichikawa} \\
$^{64}$Ni+$^{64}$Ni & 68.49 & 0.63 & 1.208 & 104.5 & 0.68 & 1.205 & \cite{Ichikawa} \\
$^{24}$Mg+$^{30}$Si & 50.52 & 0.63 & 1.172 (1.20) & 104.5 & 0.68 & 1.190 &\cite{Ichikawa} \\
$^{36}$S+$^{90}$Zr & 64.85 & 0.63 & 1.204 (1.24) & 100.0 & 0.97 & 1.07 & \cite{PRC2004}\\
$^{58}$Ni+$^{58}$Ni & 67.81 & 0.63 & 1.205 (1.19) & 104.5 & 0.68 & 1.180 & \cite{Ichikawa} \\
$^{40}$Ca+$^{40}$Ca & 59.09 & 0.63 & 1.191 (1.24) & 104.5 & 0.68 & 1.191 & \cite{Ichikawa} \\
\hline
\end{tabular}
\end{center}
\end{table}

\begin{table}
  \centering
  \caption{Comparison of the interaction barrier height $B_{int}$ for different systems between the present empirical model and Bass model. The reactions are listed in order of the increasing value of the $z$ parameter.}
  \begin{tabular}{p{3cm} p{2.5cm} p{1.5cm} p{1cm} }
    \hline
    \hline
System & $z$ &$Bass$ &$Present$     \\  \hline \\
$^{32}$S+$^{24}$Mg&$31.69$ &$25.72$ & $24.59$\\
$^{32}$S+$^{27}$Al &$33.69$ &$27.71$ & $26.97$\\
$^{18}$O+$^{64}$Ni&$33.83$ &$28.37$ & $27.15$\\
$^{18}$O+$^{62}$Ni&$34.05$ &$28.55$ & $27.40$\\
$^{18}$O+$^{60}$Ni&$34.27$ &$28.72$ & $27.67$\\
$^{16}$O+$^{64}$Ni&$34.36$ &$28.85$ & $27.77$\\
$^{18}$O+$^{58}$Ni&$34.51$ &$28.91$ & $27.94$\\
$^{16}$O+$^{62}$Ni&$34.58$ &$29.03$ & $28.03$\\
$^{16}$O+$^{60}$Ni&$34.81$ &$29.21$ & $28.30$\\
$^{18}$O+$^{65}$Cu&$34.93$ &$29.47$ & $28.44$\\
$^{18}$O+$^{58}$Ni&$35.05$ &$29.39$ & $28.58$\\
$^{18}$O+$^{63}$Cu&$35.15$ &$29.64$ & $28.70$\\
$^{35}$Cl+$^{27}$Al&$35.24$ &$29.28$ & $28.81$\\
$^{16}$O+$^{65}$Cu&$35.47$ &$29.95$ & $29.08$\\
$^{18}$O+$^{70}$Zn&$35.59$ &$30.21$ & $29.23$\\
$^{16}$O+$^{63}$Cu&$35.69$ &$30.13$ & $29.35$\\
$^{18}$O+$^{68}$Zn&$35.81$ &$30.30$ & $29.48$\\
$^{18}$O+$^{66}$Zn&$36.02$ &$30.55$ & $29.73$\\
$^{16}$O+$^{70}$Zn&$36.14$ &$30.71$ & $29.86$\\
$^{18}$O+$^{64}$Zn&$36.25$ &$30.73$ & $29.99$\\
$^{16}$O+$^{68}$Zn&$36.36$ &$30.89$ & $30.12$\\
$^{16}$O+$^{66}$Zn&$36.58$ &$31.05$ & $30.38$\\
$^{16}$O+$^{64}$Zn&$36.81$ &$31.23$ & $30.66$\\
$^{12}$C+$^{152}$Sm&$48.78$ &$44.68$ & $44.50$\\
$^{35}$Cl+$^{48}$Ti&$54.16$ &$48.39$ & $50.61$\\
$^{16}$O+$^{134}$Ba&$58.66$ &$54.44$ & $55.66$\\
$^{16}$O+$^{150}$Nd&$61.29$ &$57.41$ & $58.57$\\
$^{16}$O+$^{148}$Nd&$61.46$ &$57.57$ & $58.78$\\
$^{35}$Cl+$^{64}$Ni&$65.46$ &$60.18$ & $63.20$\\
$^{35}$Cl+$^{62}$Ni&$65.85$ &$60.49$ & $63.62$\\
$^{35}$Cl+$^{60}$Ni&$66.24$ &$60.81$ & $64.06$\\
$^{35}$Cl+$^{58}$Ni&$66.65$ &$61.13$ & $64.50$\\
$^{35}$Cl+$^{90}$Zr&$87.71$ &$83.57$ & $87.24$\\
$^{40}$Ar+$^{110}$Pd&$100.84$ &$98.01$ & $101.04$\\
$^{35}$Cl+$^{124}$Sn&$102.93$ &$100.38$ & $103.23$\\
$^{35}$Cl+$^{116}$Sn&$104.31$ &$101.52$ & $104.67$\\
$^{40}$Ar+$^{197}$Au&$153.92$ &$156.45$ & $155.61$\\
$^{40}$Ar+$^{208}$Pb&$157.95$ &$161.08$ & $159.75$\\
$^{54}$Cr+$^{207}$Pb&$202.99$ &$210.07$ & $206.98$\\
$^{52}$Cr+$^{208}$Pb&$203.78$ &$210.79$ & $207.83$\\
\hline
\hline
\end{tabular}
  \end{table}
  \section{Discussion}
  We have constructed the empirical formulae for the fusion and interaction barrier heights using the experimental results available in the literature as mentioned above. The fusion barrier heights $B_{fu}$ and fusion barrier radius $R_B$ can be obtained from the equation (\ref{eq5}) and (\ref{eq11}), respectively. The $B_{fu}$ values obtained from various theoretical models for different systems, which are not used to formulate the present model, are compared with the experimental results in table III and Fig. 2. Similar comparisons have also been done for $R_B$ in table IV and Fig. 3.
  To check which model gives the best agreement with the experimental results, the differences of both $B_{fu}$ and  $R_B$ values between the experiments and models have been plotted as a function of $z$ also in Fig. 2 and 3, respectively. Further, we obtained the sum of the squared residuals ($SSR= \displaystyle\sum_{i=1}^{n} e_i^{2}$), where $e_i$ is the $i^{th}$ residual or difference and n is the number of data points, and mean squared error ($\sigma_\epsilon^2 = \frac{SSR}{n-2}$) to find the minimum mean error ($\sigma_\epsilon $). This analysis suggests that the CW model is the best and BW model is the second best for  $B_{fu}$. Whereas the BW model is the best and AW the second best for  $R_B$. Further, the Woods-Saxon potential (equation (\ref{eq23})) is more sensitive to the $R_B$ ($R_{pt}$)  than $V_0$ and $a$ parameters. Hence, the BW model can be taken as the best model.  The BW model parameters for several reactions are compared with those obtained by others in Table V. Here, an important point is to note that though the current model is made out of the experimental results, still it is not found to be the best among the eight models considered here. It happens because of the fact that there are very few data exist for $z > 150$.\\
 
  To examine whether the BW model provides good Woods-Saxon potential, we plan to make use of the  parameters $V_0$, $r_0$ (=$R_c$), and $a_0$ 
  from this model as required by the CCFULL calculations \cite{K.Hagino} to obtain the total fusion cross sections and compare them with the experimental results. However, the radius parameters used in the BW model does not include $r_0$. To introduce it there we rewrite equation (\ref{eq23}) as follows:
  \begin{equation}
      R_{pt} = R_p + R_t + 0.29 = r_0 (A_p^\frac{1}{3}+A_t^\frac{1}{3})
      \label{eq50}
  \end{equation}
   Here $R_p$ and $R_t$ are taken from equation (\ref{eq25}).
   
  We have used such Woods-Saxon potential parameters for many reactions in the CCFULL calculations without taking the coupling between the relative motion of colliding nuclei and the intrinsic degrees of freedom into account and compared with the experimental total fusion cross sections. Most of them give very good agreements at the beam energies larger than the fusion barriers as shown for two systems $^{19}$F+$^{181}$Ta and $^{58}$Ni+$^{54}$Fe in Fig.5. However, sometimes certain departure has also been found for example $^{40}$Ca+$^{40}$Ca and $^{58}$Ni+$^{58}$Ni systems as shown in Fig.6. Note that the BW model parameters underestimate the measured cross sections for $^{40}$Ca+$^{40}$Ca and overestimates for $^{58}$Ni+$^{58}$Ni reaction. A small variation of the value of $r_0$ turns the agreement good. In the case of $^{40}$Ca+$^{40}$Ca, value of $r_0$ has to be increased from 1.19 to 1.24 fm and for $^{58}$Ni+$^{58}$Ni, the value of $r_0$ has to be decreased from 1.205 to 1.19 fm. A comparison of the Woods-Saxon potential parameters with other works for twelve reactions is shown in Table VI. Here, the asymmetric systems show better agreement than the symmetric systems. One out five asymmetric systems and four out of seven symmetric systems are not in agreement. One noticeable fact can be seen here that some of the reactions show good agreement with the measured total fusion cross sections even till the low energy side, for example, $^{19}$F+$^{181}$Ta and $^{40}$Ca+$^{40}$Ca. It means these reactions do not need any coupling effects into account to increase the cross sections anymore. Whereas, the total fusion cross sections with the potential parameters used in \cite{Moin} require the coupling effects to have agreements with the measured data. On the other hand, some reactions such as $^{58}$Ni+$^{58}$Ni and $^{58}$Ni+$^{54}$Fe require the coupling effects as usual \cite{Ichikawa}. One point may be worth noting that sometimes the CCFULL calculation does not reproduce the measured fusion cross sections and the measurement have been altered by introducing an “efficiency factor” $\epsilon$ so as to find the agreement \cite{PRC2004}. In contrast, we notice a little variation of the $r_0$ parameter gives good agreement.      \\
  
  
  Next, the interaction barrier heights can be obtained from equation (\ref{eq6}) to compare with the experiments and models. However, the measurements are very limited and all the experimental values have been used to construct the present interaction barrier model. Hence, the current values have only been compared with the Bass potential model predictions in table VI and Fig.4. Differences of the current model values from the Bass model predictions shown in the figure suggest that the present model are quite lower than the  Bass potential model predictions throughout the z values except for z$\approx$ 100. 
  
  \section{Conclusion}
  In this paper, using the experimental values available in the literature, empirical formulae for fusion and interaction barrier heights as well as barrier radii are introduced. The present study is restricted to the fusion and interaction reactions in the regime $8\leq z \leq278$ and $59\leq z \leq313$, respectively. We have carried out a comparative study of the fusion barrier as well as barrier radius between present empirical formula and various empirical and semi-empirical models along with experimental results. According to a thorough comparison with the experimental values, it is found that the Broglia and Winther model gives reasonable barrier heights in comparison with the present empirical model and various empirical and semi-empirical models. Further, to examine its predictability, the Broglia and Winther model parameters are used to obtain the total fusion cross sections and compared with the experimental values. The comparison shows good agreement at the energies above the fusion barriers, but below the barriers the predictions for some reactions show a departure from the experimental results. 
  
  Similarly, current interaction barrier heights are compared with only model available the Bass potential model predictions, because of the scarcity of measurements and other model predictions. Experiments are very essential in the lower z range. Whatsoever, the comparison shows that the present model predictions are much lower than the  Bass potential model values. We believe the current model can be used as a guideline for estimating the interaction barrier heights for the future measurements even beyond the working range of the model by extrapolation.\\ 
  
  \textbf{Acknowledgements:}
  We would like to acknowledge the illuminating discussions with Subir Nath, Ambar Chatterjee, B.R. Behra and S. Kailas. 
  
  \clearpage
\bibliographystyle{plainnat}

\end{document}